\documentclass[preprint,12pt,sort&compress]{elsarticle}

\usepackage{amssymb}
\usepackage{graphicx}
\usepackage{bm}
\usepackage{xcolor}
\usepackage{amsmath}
\usepackage{indentfirst}
\usepackage[T1]{fontenc}
\usepackage[colorlinks,linkcolor=blue]{hyperref}

\journal{Physics of the Dark Universe}

\begin{document}
\begin{frontmatter}

\title{Induced cosmological anisotropy by a gauge-gravity interaction}

\author[label1,label2]{Bum-Hoon Lee}
\ead{bhl@sogang.ac.kr}
\author[label1,label2]{Hocheol Lee}
\ead{insaying@sogang.ac.kr}
\author[label1]{Wonwoo Lee}
\ead{warrior@sogang.ac.kr}
\author[label3]{Nils A. Nilsson}
\ead{nils.nilsson@obspm.fr}
\author[label4]{Somyadip Thakur}
\ead{somyadip@sogang.ac.kr}
\affiliation[label1]{organization={Center for Quantum Spacetime, Sogang University},
             city={Seoul},
             postcode={04107},
             country={Republic of Korea}}

\affiliation[label2]{organization={Department of Physics, Sogang University},
             city={Seoul},
             postcode={04107},
             country={Republic of Korea}}

\affiliation[label3]{organization={SYRTE, Observatoire de Paris, Universit\'e PSL, CNRS, Sorbonne Universit\'e, LNE},
            addressline={61 avenue de l'Observatoire}, 
            city={Paris},
            postcode={75014},
            country={France}}

\affiliation[label4]{organization={Department of Physics, Hanyang University},
            city={Seoul},
            postcode={04763}, 
            country={Republic of Korea}}
            
\begin{abstract}
We present a simple model which generates cosmological anisotropies on top of standard FLRW geometry. This is in some sense reminiscent of the mean field approximation, where the mean field cosmological model under consideration would be the standard FLRW, and the anisotropy is a small perturbative correction on top of it. Using a supergravity-inspired model, we confirm that the stable fixed point of our model corresponds to standard FLRW cosmology. We use a Bianchi VII$_0$-type model supplemented with a scalar and $U(1)$ gauge fields, and we show that the anisotropies of the geometry are generated by the non-trivial interaction between the gravity sector and the $U(1)$ gauge sector. Studying the attractor flow, we show that the anisotropies are present at early times (high redshift) and decay asymptotically to an FLRW attractor fixed point. With such a mechanism, observations of non-isotropy are not contradictory to FLRW geometry or indeed the $\Lambda$CDM model. Such models could in principle shed some insights on the present cosmological tensions.
\end{abstract}


\end{frontmatter}

\section{Introduction} \label{intro}
One of the most successful cosmological models based on General relativity is the base Lambda Cold Dark Matter ($\Lambda$CDM) model.  This tremendously well established model of cosmology assumes a flat universe, cold dark matter
(CDM) and a positive cosmological constant, and is the simplest cosmological model which is fairly in good agreement with current observations. As the current de-facto standard model of cosmology, the spacetime geometry in $\Lambda$CDM is that of the homogeneous and isotropic Friedmann-Lema\^{i}tre-Robertson-Walker (FLRW), where the only inhomogeneities allowed are those of small perturbations, which are actually the sources of some of the most important cosmological observables. Observations of the Cosmic Microwave Background (CMB) \cite{Hu:2001bc}, Baryon Acoustic Oscillations (BAO) \cite{Weinberg:2013agg}, and Large Scale Structure \cite{Bernardeau:2001qr} have for a long time been satisfactory proof that the Universe is evolving very closely along the predictions of the $\Lambda$CDM model.

Although the standard cosmological model has been a resounding success, there exist several problems which emerge when confronting the model with data. One of the most topical is the {\it Hubble tension}, the discrepancy between the value of the Hubble constant $H_0$ when measured in the local Universe versus with CMB observations, a tension which is currently reported at $5\sigma$ \cite{Riess:2021jrx}. This is just one of a slew of ``cosmic tensions'' persistent within the $\Lambda$CDM paradigm, an overview of which can be found in \cite{Abdalla:2022yfr}. These cosmic tensions are not the only threats to $\Lambda$CDM: different types of anomalous anisotropies have been reported both in the early and late Universe, such as quadrupole-octopole alignment in the CMB \cite{deOliveira-Costa:2003utu,Schwarz:2004gk}, anomalous bulk flow \cite{Howlett:2022len,Migkas:2021zdo,Aluri:2022hzs}\footnote{For an extensive review, see \cite{Aluri:2022hzs}.}, radio-galaxy dipoles \cite{Schwarz:2004gk,Dolfi:2019wye}, and possible variations in the fine-structure constant \cite{King:2012id}. Also, recent hints of cosmic birefringence, the rotation of the polarisation plane of CMB photons, were reported at over $3\sigma$ in the \emph{Planck} EB power spectrum \cite{Minami:2020odp,Diego-Palazuelos:2022dsq,Eskilt:2022wav}. This is in sharp contrast to the $\Lambda$CDM prediction (no birefringence) and would have profound implications for fundamental physics if confirmed. It seems clear that the $\Lambda$CDM model may need to be revised.

In \cite{Watanabe:2009ct} the authors considered a model which is closely related to the case studied in our present work. Essentially, the model in this paper is a special case of the phenomenological model considered in \cite{Watanabe:2009ct}, where the coupling between the gauge field and scalar field is taken to be minimal. The authors of \cite{Watanabe:2009ct} study an inflationary scenario with a vector field coupled to an inflaton field and show that the inflationary universe is endowed with spatial anisotropy for a wide range of coupling functions $f(\phi)$, where $\phi$ is the inflaton field; importantly, the gauge-field ansatz considered in \cite{Watanabe:2009ct} is \emph{gauge inequivalent} to the gauge field considered here, and we consider the evolution of the universe without specifying the inflationary scenario. The authors in \cite{Watanabe:2009ct} focuses on the early universe evolution of the gauge fields and the inflaton fields, whereas we are more interested in the full history of the cosmological evolution after inflation.\footnote{This is in a sense complimentary to the work presented in \cite{Watanabe:2009ct}.}

In this paper, we introduce an abelian version of the chromo-natural models discussed in \cite{Komatsu:2022nvu,Ishiwata:2021yne,Maleknejad:2012fw,Maleknejad:2013npa}\footnote{In this paper we are only considering abelian gauge fields; hence, we have regular scalars fields instead of pseudo-scalars, which arises in non-abelian theories. In non-abelian theories, the scalar couples to the gauge fields as $ \phi F_{\mu\nu}\widetilde{F}^{\mu\nu}$, where  $F_{\mu\nu}\widetilde{F}^{\mu\nu}$ is a \emph{CP}-odd term, and the scalar which couples to the gauge field is therefore pseudoscalar.}. This type of model has been shown to arise naturally in $N=4$ supergravity, and has been used to study spacetime-varying couplings as discussed in \cite{Kostelecky:2002ca} and others. We begin our analysis in a very general way by using the Bianchi I spacetime before specialising to Bianchi VII$_0$ in Section~\ref{sec:eoms}, and by employing a perturbative scheme, we show that the model contains $\Lambda$CDM at the zeroth order, and that FLRW geometry is a stable point in the attractor flow. As such, there is no contradiction between the observed cosmological tensions and anisotropies and the $\Lambda$CDM model. 

This paper is organized as follows: in Section \ref{sec:GA} we introduce the model and the theoretical details; in Section \ref{sec:eoms} we discuss the covariant equations of motion and their perturbative expansions; Section \ref{sec:numsols} contains the numerical solutions, where we also present our main results; in Section \ref{darkenergyeos} we present the the dark energy equation of state generated by the gauge field and anisotropies; in Section \ref{sec:const} we compare the model to $\Lambda$CDM using low-redshift data, and we conclude in Section \ref{sec:disc}. \ref{app:biagen} contains a short treatment of the general Bianchi classification; in  \ref{klsymgf} we present the Killing symmetry of 1-form fields and the 2-form fluxes;  \ref{app:metggc} and \ref{app:const} contains the metric gauge choice and our procedure for generating initial conditions, respectively. Finally, we present the relevant Einstein equations and the perturbative expansions in \ref{app:pertexp}.

We use $c=\hbar=\kappa=8\pi G = 1$ and the metric signature $(-+++)$ throughout the paper. When studying the behaviour of the model, we focus on the time after recombination, i.e. redshift $z<1100$ and thus focus on the matter and $\Lambda$ dominated eras.

\section{Gauge-Axion model} \label{sec:GA}
In this section, we focus on the bosonic part of a supergravity-inspired model with the action
\begin{equation} \label{action1}
\begin{aligned}
    S = \int &d^4x \sqrt{-g}\Big[\frac{R-2\Lambda}{2\kappa}-\frac{1}{2}\nabla_\mu \phi \nabla^\mu \phi \\& - V(\phi) - \frac{1}{4}F_{\mu\nu}F^{\mu\nu}-\frac{\Theta \phi}{4}F_{\mu\nu}\widetilde{F}^{\mu\nu}+ \mathcal{L}_{\rm PF}\Big] ,
\end{aligned}
\end{equation}
where $\kappa=8\pi G$ (which we set to unity from now on), $R$ is the  Ricci scalar, $\Lambda$ is the cosmological constant, $\phi$ is the pseudoscalar axion field, $\Theta$ is the axion decay constant, and  $\mathcal{L}_{\rm PF}$ is the canonical Lagrange density for a perfect fluid containing baryonic matter, dark matter, and radiation. Here, $F_{\mu\nu}=2 \partial_{[\mu}A_{\nu]}$ is the field-strength tensor for the gauge field $A_\mu$ and $\widetilde{F}^{\mu\nu}=\tfrac{1}{2}\epsilon^{\mu\nu\alpha\beta}F_{\alpha\beta}$ is its dual where $\epsilon^{\mu\nu\alpha\beta}$ is Levi-Civita tensor. The new field $\phi$ can be thought of as a candidate for axionic dark matter and/or dark energy. The gauge-axion Lagrangian considered in this work is very general, which can encompass a very general class of Bianchi models; viz, Bianchi type I. In what follows, we consider the gauge field as some dark sector component, but since it is massless, it may in principle be thought of as the photon; in this case, our solutions will be further constrained by e.g. primordial magnetic fields \cite{Subramanian:2015lua}.

We note here that a stringent supergravity model would not allow us to have any explicit cosmological constant term in the action. However for the present paper where we mostly study an effective cosmological model, such constraints coming from supergravity can be relaxed and we present our action with explicit cosmological constant term.

In the rest of the paper we will mostly focus on the abelian $U(1)$ gauge field $A_\mu$, which together with the ansatz chosen makes all contributions from the symmetry-breaking term ($\propto\Theta$) vanish\footnote{In the $U(1)$ limit, the scalar and gauge fields are minimally coupled to the gravitational sector) By solving the system of coupled equations we obtain the backreacted solutions for all fields.}. We note here that for the most general gauge field and metric ansatz, i.e full dependence on the time and the spatial coordinates, the symmetry-breaking term does not vanish and has non-trivial contributions which we defer for future study. The model that we use in the present analysis can also be considered as minimally coupled Quintessence with electromagnetic fields \cite{Copeland:2003cv,Brax:2005uf,Panda:2010uq,Ibe:2018ffn}. In minimally coupled Quintessence models the Quintessence (scalar) field couples to the Maxwell term\footnote{For details, see Eq.~(2.9) in \cite{Copeland:2003cv}.}, which is in contrast to the gauge-axion model where the pseudoscalar axion couples to the $CP$-violating $\Theta$ term. 
 
It is also worthwhile to note that our analysis can be extended to non-abelian sectors, viz. $SU(2)$ or $SU(3)$ gauge groups \cite{Berghaus:2019cls,Choi:2022nlt,Papageorgiou:2022prc}, which, when coupled to the axion field would encode a QCD axion, which is among one of the most compelling candidates for physics beyond the standard model (BSM). This axion solves the strong $CP$ problem \cite{Peccei:1977hh,Weinberg:1977ma} and is potentially a natural candidate for cold dark matter \cite{Preskill:1982cy,Dine:1982ah}. In string theory, a similar spectrum of particles dubbed axion-like particles (ALPs) can be identified as ultralight dark matter with a broad mass range and interesting cosmological consequences \cite{Svrcek:2006yi,Arvanitaki:2009fg,Grin:2019mub}. In general, the abundance of axion-like dark matter is determined by the axion mass term and the coupling of the axion to the gauge sector, i.e the decay constant, which depends on the cosmological epoch when the Peccei-Quinn (PQ) symmetry breaking takes place \cite{Arias:2012az,Kawasaki:2014sqa}. In the \emph{non-abelian} case, the axion (pseudo-scalar) and gauge term coupling do not vanish and is being contributed by the Chern-Simons type terms. In this case the Einstein equations will have one extra term proportional to the structure constants $f^a_{~bc}$ (which vanish in the $U(1)$ limit).

The equations of motion derived from Eq.~\eqref{action1} are given below.

\noindent\underline{\it{The Einstein equations}}
\begin{equation} \label{eq:Eeqs}
    R_{\mu\nu}-\frac{1}{2}g_{\mu\nu}R +\Lambda g_{\mu\nu} = \tilde{T}_{\mu\nu}^{\rm PF}+T_{\mu\nu}^{\rm AN}
\end{equation}
where we add the stress-energy tensor for a perfect fluid, $\tilde{T}_{\mu\nu}^{\rm PF}$. We have simplified Eq.~\eqref{eq:Eeqs} by including the deviation from the base $\Lambda$CDM in $T_{\mu\nu}^{\rm AN}$, which we call the anisotropic stress-energy tensor; it takes the form
\begin{equation} \label{anstr}
\begin{aligned}
      T_{\mu\nu}^{\rm AN} = \nabla_\mu\phi\nabla_\nu\phi-&\frac{1}{2}g_{\mu\nu}\nabla_\alpha\phi\nabla^\alpha\phi-g_{\mu\nu}V(\phi)-\frac{1}{4}g_{\mu\nu}F_{\alpha\beta}F^{\alpha\beta}+F_{\mu}^{~\alpha}F_{\nu\alpha}.
\end{aligned}
\end{equation}
\underline{\it{Equations of motion for $\phi$ and $A_\mu$}}
\begin{eqnarray}
    0 \;&=&\; \Box\phi - V^\prime(\phi) - \tfrac{\Theta}{2}\epsilon^{\alpha\beta\delta\rho}\nabla_\beta A_\alpha \nabla_\rho A_\delta,
    \label{eq:Phieq}
    \\
	0 \;&=&\; \Box A_\mu - \nabla^\alpha\nabla_\mu A_\alpha - \Theta \epsilon_{\alpha\beta\delta\mu}\left(\nabla^\beta A^\alpha\nabla^\delta\phi+\phi\nabla^\delta \nabla^\beta A^\alpha \right). \nonumber
    \\
    \label{eq:Aeq}
\end{eqnarray}

We choose as our starting point the Bianchi I metric, which we parametrize as
\begin{equation} \label{eq:metricB1}
    ds^2 = -dt^2 + e^{2\alpha(t)}\left(e^{2\beta_1(t)}dx_1^2 + e^{2\beta_2(t)}dx_2^2 + e^{2\beta_3(t)}dx_3^2\right),
\end{equation}
where $\alpha(t)$ and $\beta_i(t)$ are the isotropic and anisotropic scale factors, respectively (for details, see \ref{app:biagen}). The factor two in the exponentials has been introduced so that the isotropic scale factor matches its FLRW equivalent, i.e. $a(t)=\text{exp}(\alpha(t))$, and $\dot{a}/a = \dot{\alpha}$. We also adopt the temporal gauge for the gauge fields and write
\begin{equation} \label{gfan}
A_\mu=
    \begin{cases}
        0, & \mu=0,\\
        A_i, & \mu=i.
    \end{cases}
\end{equation}
In \ref{klsymgf} we explicitly show that the 1-form gauge field is invariant under the Killing symmetry of the metric \eqref{eq:metricB1}, which allows us to expand the 1-form field as follows
\begin{equation}
    A_i=e_i \psi_i,
\end{equation}
where $e_i$ are the spatial triads, which take the following form\footnote{$g_{\mu\nu}=\eta_{ij}e^i_\mu e^j_\nu$}, ($\delta_i$ is the Kronecker delta)
\begin{equation}
    e_i=e^{(\alpha+\beta_i)}\delta_i.
\end{equation}
With the Bianchi I metric (\ref{eq:metricB1}) with $\mathbb{R}^3$ symmetry, we can write the gauge field as
\begin{equation} \label{eq:gaugef}
    A_i=\left(e^{\alpha+\beta_1}\psi_1,e^{\alpha+\beta_2}\psi_2,e^{\alpha+\beta_3}\psi_3\right),
\end{equation}
which allows us to rewrite the 1-form fields in terms of some scalar functions which we call $\psi_i(t),\alpha(t)$ and $\beta_i(t)$. In the following section we proceed by writing the most general coupled differential equations for the metric ansatz \eqref{eq:metricB1} and the 1-form fields \eqref{gfan}. In the rest of the paper we will focus only on the $U(1)$ 1-form field strength, and it can be shown that the symmetry-breaking term proportional to $\Theta$ vanishes identically for the abelian sector. The most general solution for non-abelian 1-form field strength will be discussed in the forthcoming paper~\cite{inprogress}.

\section{Equations of motion and their solutions} \label{sec:eoms}
We substitute the metric (\ref{eq:metricB1}) into the equations of motion (\ref{eq:Eeqs}), (\ref{eq:Phieq}), and (\ref{eq:Aeq}), and explicitly write out the results for each index value; after some simplification, we can write the scalar-field equation as
\begin{align} \label{eq:11}
    0 = \ddot{\phi} +\dot{\phi}\left(3\dot{\alpha}+\sum_k\dot{\beta}_k\right) + V^\prime(\phi).
\end{align}
With our gauge choice (temporal gauge), the temporal component of the gauge-field equation vanishes, and we can write the spatial components as  
\begin{equation} \label{eq:12}
\begin{aligned}
    0 =\ddot{\psi}_i + \Big[ \dot{\psi}_i + &\psi_i \left( \dot{\alpha} + \dot{\beta}_i \right) \Big] \left (3 \dot{\alpha} + \sum_{n=1}^3 \dot{\beta}_n \right) + \psi_i \left[ \ddot{\alpha} + \ddot{\beta}_i - \left( \dot{\alpha} + \dot{\beta}_i \right)^2 \right] .
\end{aligned}
\end{equation}
We write out all the components of the Einstein equations \eqref{eq:Eeqs} in a similar manner; these are somewhat lengthy, and we show the first Friedmann equation ($\mu=\nu=0$ component) here (the rest can be found in \ref{app:pertexp})
\begin{equation} \label{eq:13}
\begin{aligned}
     &\tilde{T}_{00}^{\rm PF}+\Lambda = 3\dot{\alpha}^2+2\dot{\alpha}\sum_{n=1}^3\dot{\beta}_n+\dot{\beta}_1\dot{\beta}_2+\dot{\beta}_2\dot{\beta}_3+\dot{\beta}_3\dot{\beta}_1 \\&\quad \quad \quad-\frac{1}{2}\sum_{n=1}^3\left[\dot{\psi}_n+\psi_n\left(\dot{\alpha}+\dot{\beta}_n\right)\right]^2-\frac{1}{2}\dot{\phi}^2-V(\phi).
\end{aligned}
\end{equation}
In the rest of this paper we incorporate the contribution from the cosmological constant $\Lambda$ into the stress-energy tensor for the perfect fluid as follows
$$T_{\mu\nu}^{\rm PF}=\tilde{T}_{\mu\nu}^{\rm PF}-\Lambda g_{\mu\nu},$$
and we work only with $T_{\mu\nu}^{\rm PF}$ (without tilde) from now on.

The stress-energy tensor $T_{\mu\nu}^{\rm PF}$ for the perfect fluid is given by\footnote{The stress-energy tensor is given by $$T_{\mu \nu}^{\rm PF}=(\rho+p)u_\mu u_\nu+p g_{\mu \nu},$$ for a boosted fluid. In this paper we consider a fluid four velocity given by$$u_\mu=(1,0,0,0),$$ with the normalization $u \cdot u=-1.$ Note that the velocity field does not receive any corrections from the non-trivial metric evolution.}
\begin{equation}
    T_{\mu\nu}^{\rm PF} = \begin{pmatrix}
    \rho & 0 & 0 & 0 \\
    0  & & & \\
    0 &  & g_{ij}p &  \\
    0 & & &
\end{pmatrix},
\end{equation}
where $\rho = \sum_i \rho_i$ is the energy density, $p = \sum_i w_i \rho_i$ is the pressure, and $w$ is the equation of state parameter, which takes the values $w=-1$ for the cosmological constant, $w=1/3$ for radiation, $w=0$ for baryonic matter, and $w=-1/3$ for curvature. Taking the flat (zero spatial curvature) case, the components of $T_{\mu\nu}^{\rm PF}$ for the homogeneous and isotropic (zeroth order) limit reads as follows
\begin{equation} \label{eq:PF}
    T_{\mu\nu}^{\rm PF,(0)}=
\begin{cases}
	3H_0^2\left(\Omega_r^0 e^{-4\alpha} + \Omega_m^0 e^{-3\alpha} + \Omega_\Lambda^0\right), \; (\mu=\nu=0) \\
	\\
	3H_0^2 \left(\tfrac{1}{3}\Omega_r^0 e^{-4\alpha} - \Omega_\Lambda^0\right)e^{2\alpha},  \quad\qquad(\mu=\nu=i),
\end{cases}
\end{equation}
where we have denoted the zeroth-order part with a superscript $(0)$. The full order can be found in \ref{app:pertexp}.

To simplify the equations of motion, we rewrite the components of the gauge field \eqref{eq:gaugef} by introducing two new scalar fields, $\sigma(t)$ and $\gamma(t)$ and redefine the $\psi_i$'s as
\begin{equation}
\begin{aligned}
    \psi_1(t)&=\frac{\psi(t)}{\sigma(t)^2-\gamma(t)^2}, \\
    \psi_2(t)&=(\sigma(t)+\gamma(t))\psi(t), \\
    \psi_3(t)&=(\sigma(t)-\gamma(t))\psi(t),
\end{aligned}
\end{equation}
which will be useful when reducing the solutions to the homogeneous and isotropic (FLRW) limit\footnote{The number of degrees of freedom is the same.}. Given these redefinitions, it is easy to see that the isotropic condition is
\begin{equation}
    \dot{\beta_i}(t)=0,  \quad\psi(t)=0,\quad \sigma(t) = \pm 1, \quad \gamma(t)=0
\end{equation}

The metric as written in Eq.~(\ref{eq:metricB1}) has had its symmetries broken down to $\mathbb{R}\times\mathbb{R}\times\mathbb{R}$, which is equivalent to the Bianchi I spacetime; in order to restore $SO(2)\times\mathbb{R}$ (or $\mathbb{R}^2\times \mathbb{R}$), we need to choose 
$$\beta_2(t)=\beta_3(t) \quad \text{ and } \quad \gamma(t)=0,$$ which sets the components of the gauge field to $A_2=A_3$. This choice brings us to the final metric which we use in the rest of this paper as
\begin{equation}
    ds^2 = -dt^2 + e^{2\alpha(t)}\left(e^{2\beta_1(t)}dx_1^2 + e^{2\beta_2(t)}(dx_2^2 + dx_3^2)\right),
\end{equation}
which is equivalent to Bianchi VII$_0$. The symmetries of this metric encapsulates the idea that the universe has a kind of preferred direction or symmetry axis, along which the cosmic expansion evolves differently.

\subsection{Perturbative Analysis}
The equations of motion in Section \ref{sec:eoms} have now been reduced to a system of coupled second-order scalar differential equations. At the zeroth order, the universe evolves in a isotropic and homogeneous space-time, and the first order contribution of the gauge-field driven anisotropy is small. The intergalactic gauge field decays away rapidly~\cite{Zeldovich:1983cr} in the late-time evolution, and since this gauge field is driving the anisotropies, we can therefore study them perturbatively.

In order to obtain numerical solutions, we use a perturbative approach and employ the following scheme:
\begin{itemize}
    \item{Expand all scalar degrees of freedom $\zeta = \{\alpha,\beta_i,\phi,\psi,\sigma\}$ in a perturbative series around their equilibrium fixed points (homogeneous and isotropic fixed point) and retain only the linear order in perturbations  
    \begin{equation}
        \zeta(t) = \zeta^{(0)}(t) + \epsilon\,\zeta^{(1)}(t),
    \end{equation} 
    where $\epsilon$ is a book-keeping device for perturbative order.}
    \item Find the zeroth-order ($\epsilon\rightarrow 0$) solutions.
    \item Plug the zeroth-order solutions back into the equations, where they act as seed solutions for first order.
\end{itemize}
By introducing the perturbative parameter $\epsilon$, we explicitly note that the metric anisotropies are small, but we have not linearised the new metric functions $\beta_i^{(1)}$.
Following the above scheme we write out the perturbative expansions around the homogeneous and isotropic fixed points as
\begin{equation}
\begin{aligned}
    \alpha(t) &= \alpha^{(0)}(t), ~ \sigma(t) = \pm 1+\epsilon\,\sigma^{(1)}(t), \\
    \beta_1(t) &= \epsilon\,\beta_1^{(1)}(t), ~ \beta_{2}(t) = \epsilon\,\beta_2^{(1)}(t), ~\beta_3(t)=\beta_2(t),\\
    \psi(t) &= \psi^{(0)}(t)+\epsilon\,\psi^{(1)}(t), ~ \phi(t) = \phi^{(0)}(t)+\epsilon\,\phi^{(1)}(t),
\end{aligned}
\end{equation}
where we have used the remaining gauge freedom in the metric to set $\alpha^{(1)}(t)=0$~(For details, see \ref{app:metggc}). We have also set $\sigma^{(0)}(t) = 1$ and $\beta_i^{(0)}(t)=0$, since this represents the homogeneous and isotropic zeroth-order background; moreover, we set $\gamma(t)=0$ to restore the planar $SO(2)\times \mathbb{R}$ symmetry. 

The perfect fluid evolves according to the continuity equation, which in the $\Lambda$CDM case reads $\dot{\rho}+3H(1+w)\rho = 0$. This equation changes due to the present non-trivial Bianchi $VII_h$ geometry \cite{Appleby:2012as,2010PhRvD..81h1301A}. The implications and perturbative corrections to the contiuity equation and $T_{\mu\nu}^{\rm PF}$ are presented in \ref{app:pertexp}.

From now on, expressions of order $\epsilon$ will always be enclosed in square brackets.

\subsubsection{Zeroth order}
As a first consistency check, we start with the zeroth-order vacuum equations, where we set  ($T_{\mu\nu}^{\rm PF}=0$) and  $\phi^{(0)}(t) = \psi^{(0)}(t) = 0$, leaving us with a system of equations which is the flat-space vacuum. In this case, the system we need to solve is the two Friedmann equations, which read
\begin{equation}
        3(\dot{\alpha}^{(0)})^2 = 0,\quad 3(\dot{\alpha}^{(0)})^2 + 2\ddot{\alpha}^{(0)} = 0.
\end{equation}
The allowed solution of the above equation gives a constant solution for the $\alpha^{(0)}$. 
With the identification of  $\alpha^{(0)}(t)=\log{a(t)}$ gives the physical scale factor, and  which reduces to the familiar solution for a static Universe.  

Adding now a radiation term in the stress-energy tensor, the Friedmann equations read
\begin{equation}
    \begin{aligned}
        (\dot{\alpha}^{(0)})^2 &= H_0^2\Omega_r^0 e^{-4\alpha^{(0)}}, \\ 2\ddot{\alpha}^{(0)}+3(\dot{\alpha}^{(0)})^2 &= -H_0^2\Omega_r^0e^{-2\alpha^{(0)}},
    \end{aligned}
\end{equation}
which solves as
\begin{equation}
    \alpha^{(0)}(t) = \frac{1}{2}\ln{\left(2H_0\sqrt{\Omega_r^0}\right)}+\frac{1}{2}\ln t,
\end{equation}
and the corresponding scale factor reads \\$a(t)=(2H_0\sqrt{\Omega_r^0})^{1/2} \sqrt{t}$, which is consistent with standard FLRW evolution. 

We now turn our attention to the more general case when $\phi^{(0)}$ and $\psi^{(0)}$ are non-zero. Here, the dynamical variables are $\phi^{(0)}$, $\phi^{(0)}$ and $\psi^{(0)}$, and we have the following three equations for the scalar, gauge field, and Einstein parts, respectively
\begin{equation}
    \begin{aligned}
        0&=\ddot{\phi}^{(0)}+3\dot{\alpha}^{(0)}\dot{\phi}^{(0)}+V^\prime(\phi^{(0)}) \\
        0&=\ddot{\psi}^{(0)}+3 \dot{\alpha}^{(0)} \dot{\psi}^{(0)}+\left(\ddot{\alpha}^{(0)}+2(\dot{\alpha}^{(0)})^2\right)\psi^{(0)} \\
        &T_{00}^{\rm PF, (0)}= 3 (\alpha^{(0)})^2  - \frac{3}{2}\left( \dot{\psi}^{(0)} + \psi^{(0)} \dot{\alpha}^{(0)} \right)^2  - \frac{1}{2}(\phi^{(0)})^2 - V'(\phi^{(0)}) .
    \end{aligned}
\end{equation}
By examining the full set of equations in \ref{app:pertexp}, we notice that all terms containing $\sigma^{(1)}$ or $\beta_i^{(1)}$, i.e. the \emph{anisotropic} variables, are proportional to $\psi^{(0)}$ or its time derivative. This influences our choice of initial conditions in the numerical solutions: if we simply choose $\psi^{(0)}(0)=\text{const.}$ and $\dot{\psi}^{(0)}(0)=0$, we obtain a solution proportional to a constant $\psi^{(0)}$, and this can simply be gauged away. In order to obtain a meaningful solution, we therefore have to implement a non-zero $\dot{\psi}^{(0)}$ as our initial condition. A description of our method for choosing consistent initial conditions can be found in \ref{app:const}.

\subsubsection{First order}
The scalar equation \eqref{eq:11} reads
\begin{equation}
\begin{aligned}
   0 =&\ddot{\phi}^{(0)}+3 \dot{\alpha}^{(0)} \dot{\phi}^{(0)}+V'(\phi^{(0)}) +\epsilon  \Big[\left( \dot{\beta}^{(1)}_1 + 2\dot{\beta}^{(1)}_2 \right) \dot{\phi}^{(0)}+ \ddot{\phi}^{(1)} + 3 \dot{\alpha}^{(0)} \dot{\phi}^{(1)} \\&+ \phi^{(1)} V''(\phi^{(0)})\Big],
\end{aligned}
\end{equation}
where the factor 2 on $\beta_2^{(1)}$ comes from $\beta_2=\beta_3$. The expressions for the gauge field and Einstein equations are rather lengthy, and we will only display the zeroth order in this section, including the full equations in \ref{app:pertexp}. 

For the gauge field in Eq.~\eqref{eq:12}, the zeroth-order expressions are identical $\mu=1,2,3$, but the equations differ at first order, and due to the symmetries, the $\mu=2$ and $\mu=3$ components are equal. Keeping to our choice of a positive sign for $\sigma^{(0)}=+1$, all the spatial components are identical (at zeroth order), and read
\begin{align}
    0=\ddot{\psi}^{(0)}+3 \dot{\alpha}^{(0)} \dot{\psi}^{(0)}+\left( \ddot{\alpha}^{(0)} + 2 (\dot{\alpha}^{(0)})^2 \right)\psi^{(0)}+ \mathcal{O}(\epsilon).
\end{align}
The first Friedmann equation ($\mu=\nu=0$ component of the Einstein equations) read
\begin{equation}
\begin{aligned}
     T_{00}^{\rm PF, (0)}+\epsilon\,T_{00}^{\rm PF, (1)} &= 3 (\alpha^{(0)})^2  - \frac{3}{2}\left( \dot{\psi}^{(0)} + \psi^{(0)} \dot{\alpha}^{(0)} \right)^2 \\ &\quad  - \frac{1}{2}(\phi^{(0)})^2 - V'(\phi^{(0)}) + \mathcal{O}(\epsilon).
\end{aligned}
\end{equation}
The spatial diagonal components ($\mu=\nu=i$) are identical at zeroth order and read
\begin{equation}
\begin{aligned}
     &T_{11}^{\rm PF, (0)}+\epsilon\,T_{11}^{\rm PF, (1)} = -e^{2 \alpha^{(0)}} \Big[2 \ddot{\alpha}^{(0)} + 3 (\dot{\alpha}^{(0)})^2 \\ &+ \left( \dot{\psi}^{(0)} + \dot{\alpha}^{(0)} \psi^{(0)} \right)^2 + \frac{1}{2} (\dot{\phi}^{(0)})^2 - V(\phi^{(0)}) \Big]+\mathcal{O(\epsilon)}.
\end{aligned}
\end{equation}
We choose a simple $\phi^4$-type potential for $V(\phi)$ as
\begin{equation}
    V(\phi) = V_0 \phi^4,
\end{equation}
where $V_0$ is a constant, and we expand $V(\phi)$ and its derivatives; the potential reads
\begin{equation}
\begin{aligned}
    V(\phi) &= V_0(\phi^{(0)})^4 +\epsilon\left(4V_0(\phi^{(0)})^3\phi^{(1)}\right), \\
    V^\prime(\phi) &= 4V_0 (\phi^{(0)})^3 + \epsilon\left(12V_0(\phi^{(0)})^2\phi^{(1)}\right), \\
    V^{\prime\prime}(\phi) &= 12V_0(\phi^{(0)})^2 + \epsilon\left(24V_0\phi^{(0)}\phi^{(1)}\right).
\end{aligned}
\end{equation}
In order for the kinetic term to not dominate over the potential at all times, we have set the value of the constant, $V_0= 10^{-3}$ in our numerical computation.

\section{Numerical solutions} \label{sec:numsols}
We solve the full system of coupled differential equations for scalar, gauge field, and Einstein parts order-by-order and present the relevant solutions here; the full equations can be found in \ref{app:pertexp}. When generating these solutions we fix the background FLRW cosmology to the parameter set $H_0=70$ km s$^{-1}$ Mpc$^{-1}$, $\Omega_m=0.3$, $\Omega_\Lambda=0.7$.
The qualitative behaviour of these solutions indicate that the field content $\phi(t)$ and $A_\mu(t)$ have considerable contribution in the early Universe before decaying exponentially, and eventually flowing to the homogeneous and isotropic attractor fixed point, which exactly corresponds to FLRW. 
\begin{table}[h]
\begin{center}
\begin{tabular}{ccc}
    Zeroth order & & \\
    \hline\hline \\[-1mm]
    $\phi^{(0)}(t_f) =10^{-6}$ & $\psi^{(0)}(t_f) = 10^{-6}$ & $\alpha^{(0)}(t_f) = 1.6$ \\
    $\dot{\phi}^{(0)}(t_f) = -10^{-6}$ & &\\[2mm]
     First order & & \\
    \hline\hline \\[-1mm]
    $\phi^{(1)}(t_f) = 10^{-6}$ & $\psi^{(1)}(t_f) = 10^{-6}$ & $\sigma^{(1)}(t_f) = 10^{-3}$ \\
    $\dot{\phi}^{(1)}(t_f) = -10^{-6}$ & $\dot{\psi}^{(1)}(t_f) = -10^{-6}$ & $\dot{\sigma}^{(1)}(t_f) = -10^{-3}$ \\
    $\beta^{(1)}_1(t_f) = 10^{-6}$ & $\beta^{(1)}_2(t_f) = -10^{-6}$ & $\dot{\beta}^{(1)}_1(t_f) = -10^{-6}$ \\[1mm]
\end{tabular}
    \caption{Boundary conditions used in the numerical solutions, defined at $t_f = 20$ Gyr.}
\label{tab:initconds}
\end{center}
\end{table}
The initial conditions for all the variables are in general coupled, and need to satisfy the equations of motion; therefore, the conditions shown in Table~\ref{tab:initconds} are the ones we choose as ``primary'', whilst the rest are derived. In \ref{app:const} we present our method for finding the rest of the boundary conditions from the Einstein equations in a consistent way.

From the zeroth-order equations we can solve the isotropic part of the scale factor $\alpha^{(0)}$ from the zeroth-order Einstein equations. Here we have imposed boundary condition at the isotropic fixed point and solved the evolution of the Einstein equations. The evolution of the zeroth order scalar and the gauge fields, $\phi^{(0)}$ and $\psi^{(0)}$ respectively.

The second order differential equations governing the evolution of the Einstein equations, 1-form gauge fields and the scalars are roughly damped harmonic oscillators, the solutions of which contain both growing and decaying modes; however, to be consistent with observations of the late-time universe, the evolution should settle down to homogeneous and isotropic solutions, viz. FLRW universe. In order to keep consistency with the cosmic no-hair theorem (the scalar/hairy solution should decay at late times) we have imposed the boundary condition at ($t \sim 20$ Gyr); the evolution at early times is governed by the Einstein equations.

In our numerical solutions we retain the decaying solutions. \\

\begin{figure}[h]
\centering
    \includegraphics[width=0.8\textwidth]{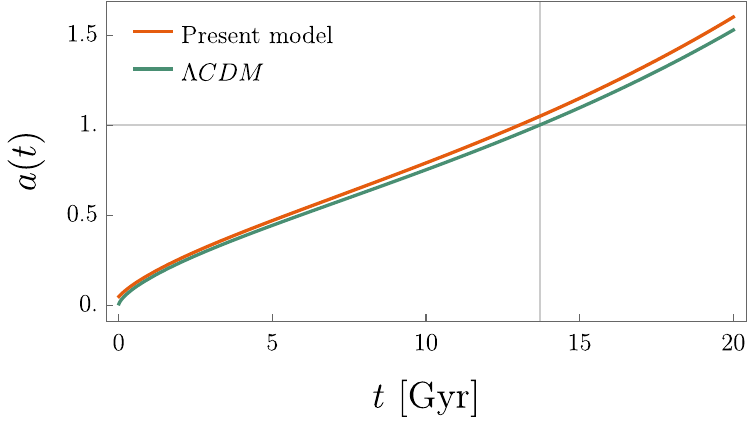}
    \caption{The isotropic scale factor $a(t)=e^{\alpha^{0}(t)}$ compared with the $\Lambda$CDM model.}
    \label{fig:alpha0}
\end{figure}

\begin{figure}[h]
\centering
    \includegraphics[width=0.8\textwidth]{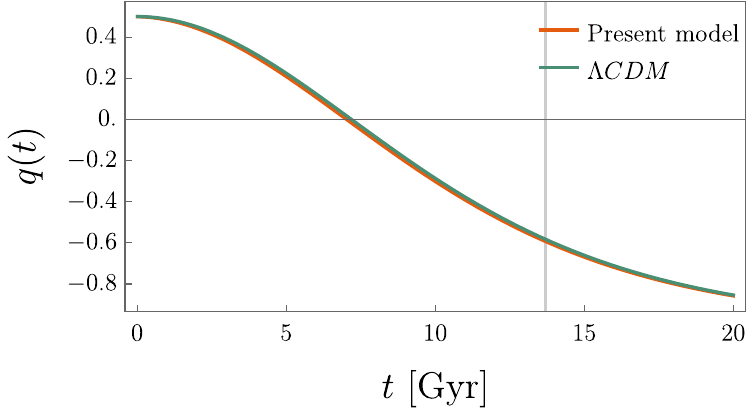}
    \caption{The deceleration parameter q compared with that of $\Lambda$CDM.}
    \label{fig:decel}
\end{figure}

\begin{figure}[h]
\centering
    \includegraphics[width=0.8\textwidth]{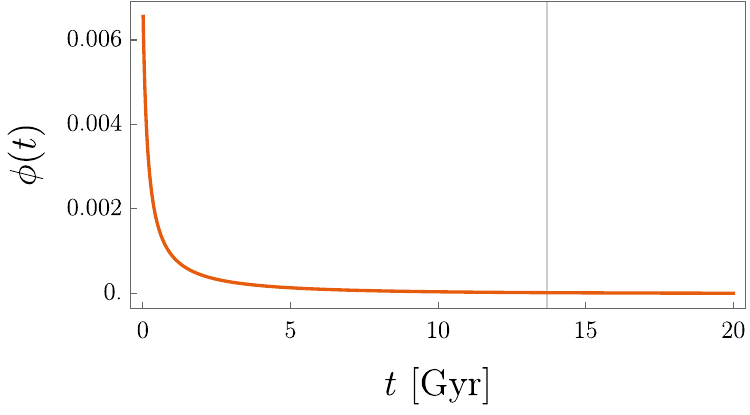}
    \caption{The behaviour of the full scalar field $\phi(t) = \phi^{(0)}+\epsilon \phi^{(1)}$.}
    \label{fig:phi}
\end{figure}

\begin{figure}[h]
\centering
    \includegraphics[width=0.8\textwidth]{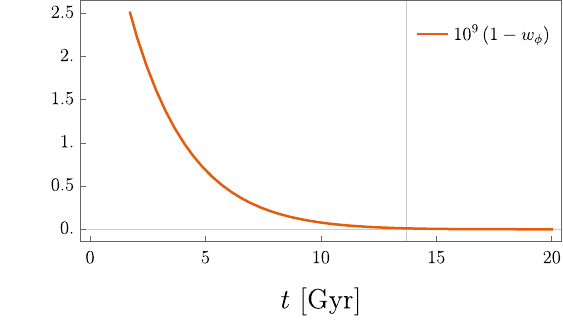}
    \caption{The behaviour of the equation of state for the scalar field $\phi$.}
    \label{fig:wphi}
\end{figure}

\begin{figure}[h]
\centering
    \includegraphics[width=0.8\textwidth]{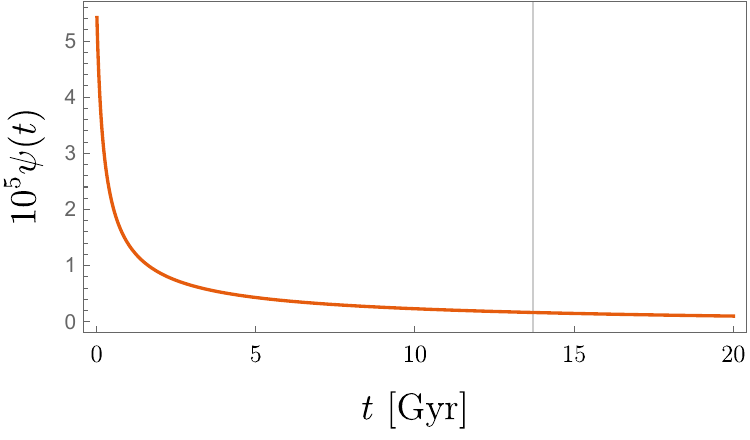}
    \caption{The behaviour of  $\psi(t) = \psi^{(0)}+\epsilon \psi^{(1)}$.}
    \label{fig:psi}
\end{figure}

\begin{figure}[h]
\centering
    \includegraphics[width=0.8\textwidth]{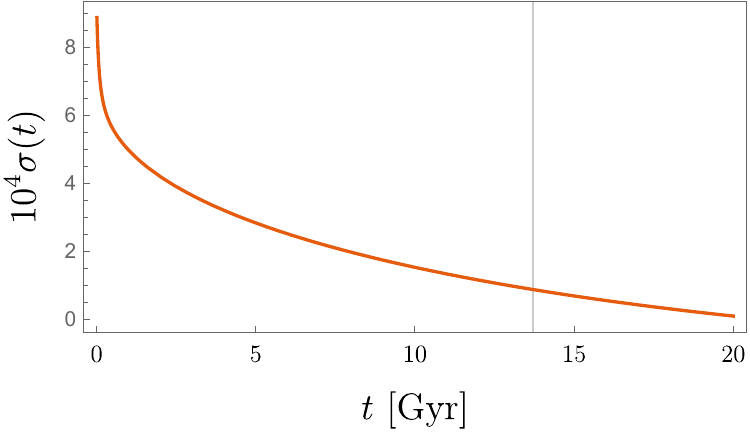}
    \caption{The behaviour of  $\sigma(t) =\epsilon \sigma^{(1)}$.}
    \label{fig:sigma}
\end{figure}

\begin{figure}[h]
\centering
    \includegraphics[width=0.8\textwidth]{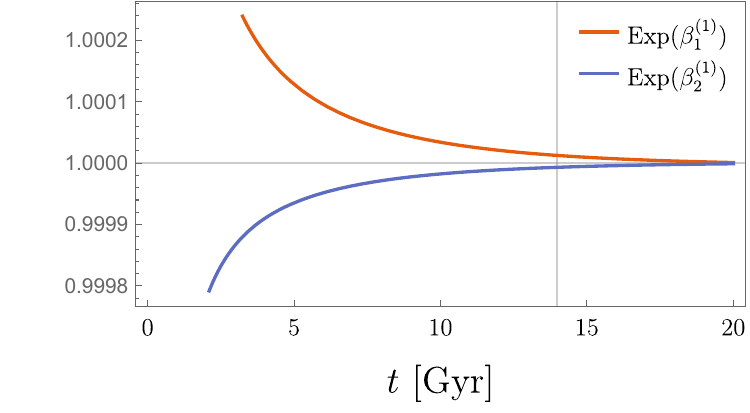}
    \caption{The anisotropic scale factors $\beta^{(1)}_1$ and $\beta^{(1)}_2$. Note that these should be further suppressed by $\epsilon$.}
    \label{fig:betas}
\end{figure}

\begin{figure}[h]
\centering
    \includegraphics[width=0.8\textwidth]{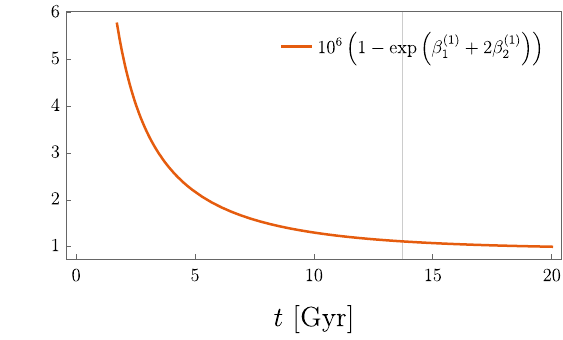}
    \caption{The total anisotropic scale factor.}
    \label{fig:betasum}
\end{figure}

\begin{figure}[h]
\centering
    \includegraphics[width=0.8\textwidth]{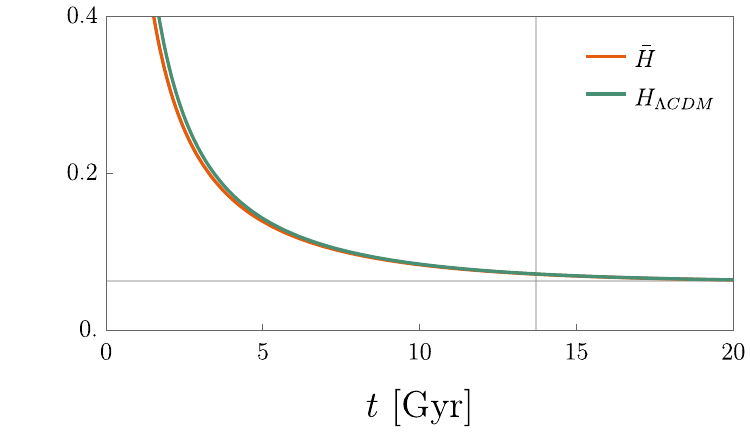}
    \caption{The average Hubble parameter in contrast to the pure $\Lambda$CDM case represented by $H_{\rm \Lambda CDM}$, which is the isotropic part of the Hubble parameter.}
    \label{fig:Haverage}
\end{figure}

\begin{figure}[h]
\centering
    \includegraphics[width=0.8\textwidth]{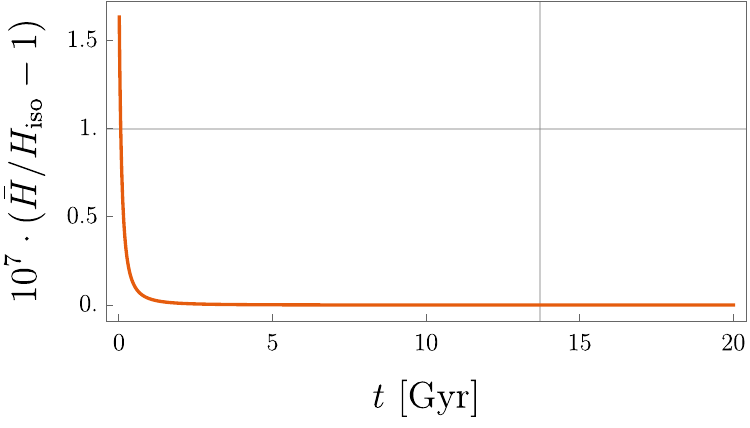}
    \caption{The average scale factor normalized by the isotropic case.}
    \label{fig:Hratio}
\end{figure}

 \underline{\it{Numerical results:}}
 \begin{itemize}
     \item In Figure~\ref{fig:alpha0} we present the solution of the isotropic scale factor. Our result at current epoch, viz. $t_0=H_0^{-1}=13.7$ Gyr, in good agreement with the results in \cite{Daniel:2009bst}. The isotropic scale factor has been plotted against the scale factor of $\Lambda$CDM (which has been normalized to unity at the present time. The deviation from the $\Lambda$CDM value can be attributed to the scalar and gauge fields in the present model under study.
     
     Next we focus on the deceleration parameter, which for $\Lambda$CDM is canonically defined in terms of the scale factor (a(t)) as
     \begin{equation}
        q(t) = - \frac{\ddot{a}(t)a(t)}{\dot{a}(t)^2}
     \end{equation}
     In Figure~\ref{fig:decel} we compare the deceleration parameter for the model under consideration with $\Lambda$CDM, and we notice that the present model has marginally faster expansion ($q$ more negative), with the difference being most pronounced between $t=3-10$ Gyr. This faster expansion is expected to play a crucial role in alleviating $H_0$ tension in this model.  
     
     \item In Figure~\ref{fig:phi} we present the solution for the scalar fields. 
     The scalar field profile starts with a non-zero divergent nature in the early universe, before rapidly decaying and finally saturating to zero at very late asymptotic times. This axion-like particle can be attributed to the scalar dark sector contributing to either dark energy (and/or dark matter). In the following section \ref{darkenergyeos} we examine the energy equation of state, which confirms our observations here. We also show the evolution of the equation of state for the scalar field $\phi$ in Figure~\ref{fig:wphi}, which can be seen to exhibit kination behaviour for most of cosmic history, only decreasing in value slightly at very early times.
     
     \item In Figure~\ref{fig:psi} and \ref{fig:sigma} we show the behaviour of the fields $\psi$ and $\sigma$, both of which take on very small values, even at early times, before flowing to the attractor fixed point asymptotically, which is consistent with our construction. Essentially there will be no residual gauge fields in the future and only residual gauge-field contributions would survive to the present epoch $\sim$ 13.7 Gyr; this is consistent with present observations.
     
     One crucial point at this juncture is to bear in mind the overall picture: the backreaction from the $U(1)$ gauge fields are  generating the anisotropies in the early Universe, and the anisotropies settle down to their fixed-point values as the gauge field saturates to the attractor fixed points.
     
     \item
     The zeroth-order solutions of the Friedmann equations dictate the isotropic evolution of the universe, which is the base $\Lambda$CDM; however, we notice that there is some deviation due to the residual presence of the scalar and gauge-field contributions, where 
    the contribution from the anisotropic parameters appear as perturbative corrections.
    
     The anisotropic contributions to the metric, $\beta_1$ and $\beta_2$, are suppressed by order $10^{-6}$ as compared to the isotropic scale factor, which is in agreement with the observational constraints where the anisotropy in the universe is comparatively very small as compared to the isotropic scale factor. In Figure~\ref{fig:betas} we show the evolution of the anisotropic scale factors exp$(\beta_1)$ and exp$(\beta_2)$, which flow towards the stable fixed point at late times, exactly the isotropic limit (Note that $\beta^{(1)}_1$ and $\beta^{(1)}_1$ should be further suppressed by $\epsilon$), in keeping with observational results. The apparent mirror similarity in Figure~\ref{fig:betas} is a consequence of the coupled nature of the equations of motion, where we are only able to choose three out of the four initial conditions related to the $\beta_i^{(1)}$'s (as seen in Table~\ref{tab:initconds}), and the fourth condition is then imposed for self-consistency (as shown in \ref{app:const}), which selects the depicted solutions for the anisotropic scale factors. We also present the total anisotropic scale factor, which is the exponential sum of the $\beta^{(1)}_i$'s, where we clearly see that it saturates to unity at late times, since the anisotropies decay; Figure~\ref{fig:betasum} depicts this behaviour, clearly showing the return of homegeneity and isotropy at late times.

     \item In order to quantify the evolution of the anisotropic degrees of freedom, we define the \emph{average} Hubble parameter $\bar{H}$ as follows
     \begin{equation}\label{eq:hbar}
         \bar{H}(t) = \frac{1}{3}\left(3\dot{\alpha}^{(0)}(t)+\epsilon \dot{\beta}^{(1)}_1(t)+2 \epsilon \dot{\beta}^{(1)}_2(t)\right).
     \end{equation}
     
     In Figure~\ref{fig:Haverage} and \ref{fig:Hratio} we show the full contribution of the anisotropy to the Hubble parameter compared to base $\Lambda$CDM. From these two plots we can see that the average Hubble parameter $\bar{H}$ is slightly smaller than its $\Lambda$CDM counterpart at all times, but that this difference is larger at early times. We also see that when compared to the isotropic limit of the present model (Figure~\ref{fig:Hratio}), the effects of the anisotropies are on the order of $\leq 10^{-7}$ throughout the history of the universe, though divergent as very early times\footnote{The primordial universe lies beyond the scope of this paper, since we neglect the contribution from the radiation $\Omega_r^0$, which dominates in that epoch.}. 

     Using the isotropic Hubble parameter (Eq.~(\ref{eq:hbar}) for $\beta_i^{(1)}\to 0$) we can construct the time-dependent energy densities for matter and $\Lambda$ as $\Omega_X(t)\equiv \rho_X/\rho_{\rm c}$, where $\rho_c$ is the critical density. Using these quantities we can establish the relative contributions of matter and $\Lambda$ to the total energy budget of the Universe across cosmic history. We also form the analogue of the energy densities when taking anisotropic evolution into account the average Hubble parameter and scale factor. We plot these quantities in Figure~\ref{fig:Omegaevol} (where quantities formed with the average quantities are denoted with an overbar). Due to the attractive nature of the potential, we see a generally lower values at early (late) times for matter ($\Lambda$), which causes the deviation in the deceleration parameter seen in Figure~\ref{fig:decel}.     
     
     The effects of the anisotropic variables on cosmic evolution may be important when studying the $H_0$ tension and other cosmological puzzles, but a detailed treatment of observational signatures consistent with the observational signatures of the $H_0$ tension requires some more exploration and lies beyond the scope of this paper, although we give some brief comments below.
 \end{itemize}

\begin{figure}[h]
\centering
    \includegraphics[width=0.8\textwidth]{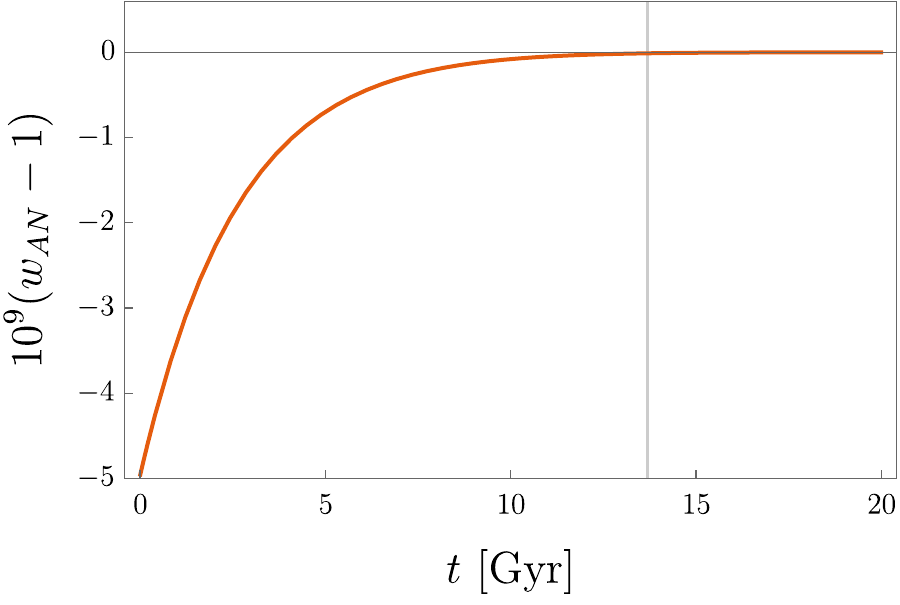}
    \caption{The dimensionless number of anisotropic equation of state.}
    \label{fig:wAN}
\end{figure}

 We end this section with some plausible implications of our axion-anisotropic cosmological model on the resolution of the present cosmological tensions. A naive observation from the solution of the average Hubble parameter from Figure~\ref{fig:Haverage} indicates that the value of Hubble parameter is lower than in the base $\Lambda$CDM model, especially at very early times. A natural question to ask at this juncture is: {\it{Can the Hubble tension be resolved in the presence of some extra degrees of freedom on top of standard FLRW cosmology?}} Let us briefly present the possibility of the model under consideration in resolving one the specific cosmological tension; viz, the $H_0$ tension. For an efficient resolution of the $H_0$ tension in the context of any effective-field theory approach, the predicted Hubble parameter should be large ($\sim 73$ km s$^{-1}$Mpc$^{-1}$) compared to the standard prediction from the astrophysical models of $\Lambda$CDM. A quick comparison of the Hubble parameter for the model under consideration and that of $\Lambda$CDM in Figure \ref{fig:Hratio} indicates that the Hubble parameter of $\Lambda$CDM should be higher; a
naive conclusion would be that the model presented in this paper not efficient in resolving the $H_0$ tension effectively. Some plausible explanations for this include
\begin{itemize}
    \item In Figure \ref{fig:wAN}, we observe that the dominant contribution to the dark energy induced by the anisotropic matter sector is controlled by the scalar fields; effectively, the kinetic terms of the scalar fields are dominant (which is why the energy equation of state saturates to unity) and the contribution of the gauge
fields are negligible. This gives a possible explanation: as the Universe starts to expand under the gravitational force, the scalar fields tries to counterbalance the expansion; thus, there is small dip in the Hubble parameter compared to $\Lambda$CDM.
    \item This is in general true for any EFT which has dominant contribution from the bosonic sectors.
\end{itemize}

In \cite{Berghaus:2019cls} the authors showed that a rolling axion coupled to a non-Abelian gauge field has the potential to provide a viable solution to the Hubble tension. The pertinent point made in \cite{Berghaus:2019cls} is that the axion fields coupled to non-abelian gauge fields provides some additional \emph{friction} term (thermal friction) to the gravity system, and thus have a potential solution to stabilize the Hubble tension. 

\begin{figure}[h]
\centering
    \includegraphics[width=0.8\textwidth]{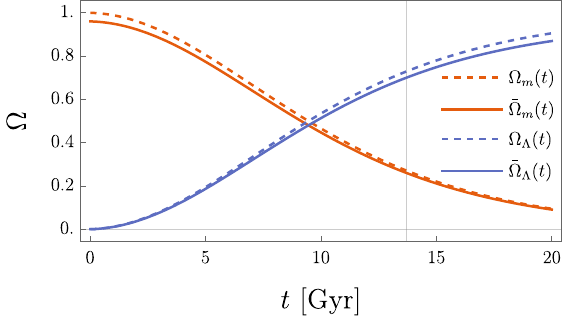}
    \caption{The evolution of the fully isotropic energy density for matter and $\Lambda$ (without bar) compared to the corresponding quantities constructed using the average scale factor and Hubble parameter (with bar).}
    \label{fig:Omegaevol}
\end{figure}

\section{Anisotropic dark energy}\label{darkenergyeos}
From our construction it is worthwhile to investigate the anisotropic contribution to the energy equation of state. We can write the anisotropic stress-energy tensor \eqref{anstr} in the standard form as 
\begin{equation}
    T_{\mu\nu}^{\rm AN} = 
\begin{pmatrix}
    \rho^{\rm AN} & 0 & 0 & 0 \\
    0  & & & \\
    0 &  & g_{ij} p^{\rm AN}_i &  \\
    0 & & &
\end{pmatrix}.
\end{equation}
In the particular case of homogeneous and isotropic cosmological models, we can assume an equation of state of the form
\begin{align}
P=\omega \rho,
\end{align}
and in the presence of anisotropic matter sources and geometry, the total pressure and the total energy density can similarly be split into isotropic and anisotropic parts 
\begin{equation}
\begin{aligned}\label{eq:rhopAN}
    \rho_t &= \left(\rho^{\rm PF}+\rho_0^{\rm AN}\right)+\epsilon \rho_1^{\rm AN},\\
    P_t &= \left( P^{\rm PF}+P_i^{\rm AN(0)}\right)+\epsilon P_i^{\rm AN(1)},
\end{aligned}
\end{equation}
from which we can determine the effective equation of state parameter $w_t$ for the cosmic fluid, as was also noted in \cite{Koivisto:2005mm,Koivisto:2008ig,Appleby:2012as,2010PhRvD..81h1301A,Guarnizo:2020pkj}. Note that we show in \ref{app:pertexp} that the perfect-fluid part also receives corrections at order $\epsilon$; these contributions are coupled to the anisotropic degrees of freedom, and we count them as part of $\rho_1^{\rm AN}$ and $P_i^{\rm AN}$.
\begin{figure*}[t]
\centering
    \includegraphics[width=0.7\textwidth]{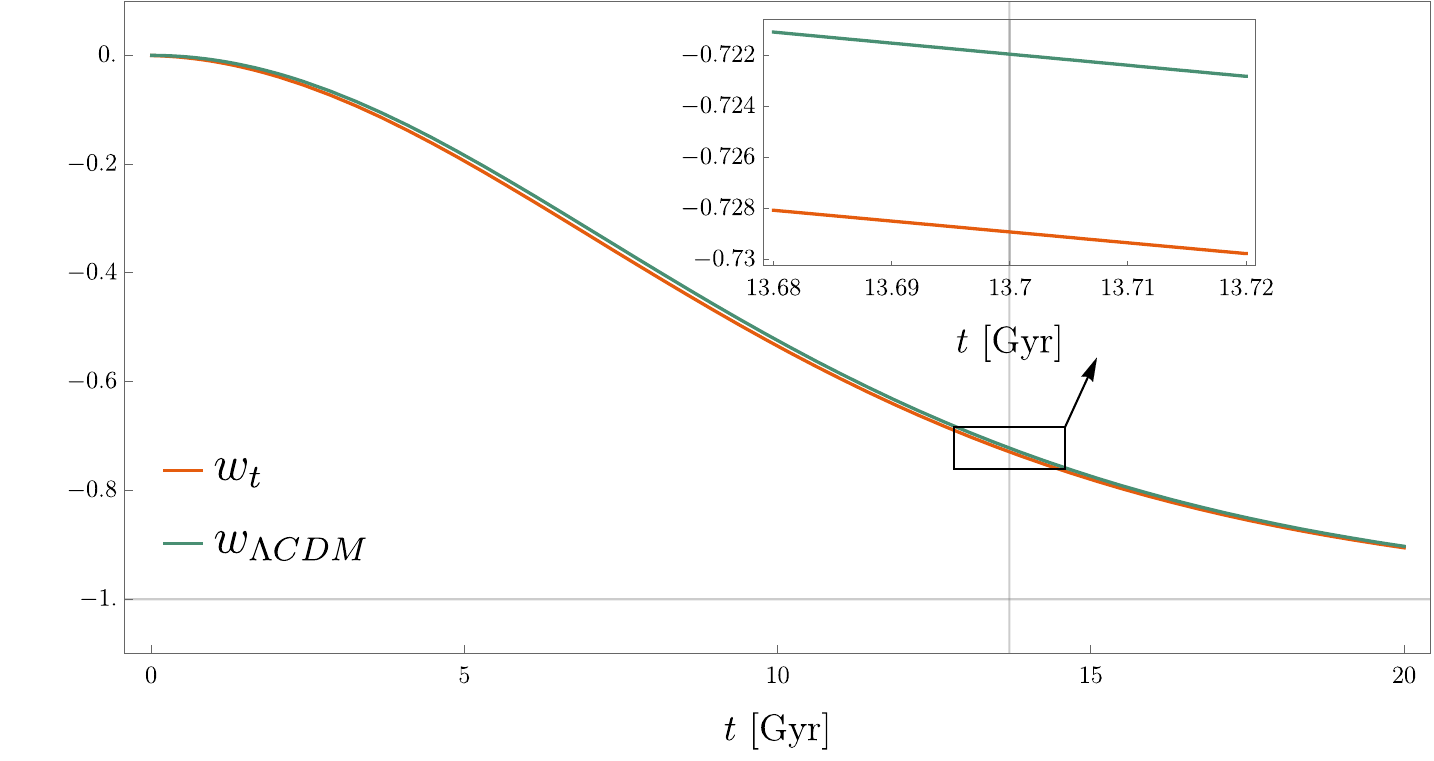}
    \caption{The behaviour of the total equation of state parameter $w_{t}$ compared to that of $\Lambda CDM$}
    \label{fig:wt}
\end{figure*}
In Figure~\ref{fig:wt} we show the evolution of $w_{\rm t}$ as a function of time, and we observe that it stays negative throughout all of cosmic history, and is close to, but always lower than, the $\Lambda$CDM model. From the point of view of the perfect fluid, the negative values of the equation of state parameter are to be expected, since we neglect the radiation term $w_{\rm r}=1/3$, and $\omega \leq$ 0 for both matter and cosmological constant. 

It is also interesting to examine the contribution to $w_t$ from the anisotropic variables. First of all, by examining the anisotropic energy density $\rho_1^{\rm AN}$ in Eq.~\eqref{eq:rhopAN} and comparing it to the perfect fluid, we see that $\rho^{\rm PF}$ dominates, and the anisotropic parts make up on the order of $10^{-5}$ of the total energy budget of the system. Moreover, when examining the equation of state for the anisotropic contribution (which we may call $w_{\rm AN}$), we see that up to a few parts in $10^8$, $w_{\rm AN}$ is a constant throughout cosmic history, with a value of $$w_{\rm AN} \approx 1.$$ This corresponds to a \emph{stiff matter} fluid, which has been studied in the context of both classical and quantum cosmology in \cite{Chavanis:2014lra} and \cite{Oliveira-Neto:2011uhf} and others. Specifically, it was found in \cite{Chavanis:2014lra} that a stiff fluid may lead to a bouncing solution of the Wheeler-de-Witt equation.

\section{Exploring the parameter space}\label{sec:const}
Given the predictions our model makes, it is interesting to compare it to some available data. In this Section, we perform a post-fit analysis\footnote{A full Markov-Chain Monte Carlo (MCMC) analysis of our model is in progress.} using late-time cosmological data at the background level (using only distance measures). In this paper, we have considered the case of vanishing radiation density ($\Omega_r^0=0$), and our model should therefore provide its best fit at low redshift; late-time data should therefore be sufficient to gain some insight of the overall fit of the model. To accomplish this, we employ a combination of two robust local-Universe datasets as described below.

We use the Pantheon+ catalogue of Type Ia supernovae (SNeIa) with SH0ES Cepheid host calibrators \cite{Scolnic:2021amr, Brout:2022vxf}, which is a set of 1701 light curves and 1550 resolved SNeIa in the redshift range $0.001 < z < 2.26$. The inclusion of Cepheids with known distances provides a robust calibration of the SneIa lightcurves and breaks the degeneracy between $H_0$ and $\Omega_m^0$. In order to compare the model with this dataset, we construct the theoretical distance modulus as 
\begin{equation}
\mu_{\rm th}(z_{\rm hel},\bm{\theta})=25+5\log(d_L(z_{\rm hel},\bm{\theta})),
\end{equation}
where $z_{\rm hel}$ is the redshift in the heliocentric frame, $\bm{\theta}$ is a vector containing the model parameters, and $d_L$ is the luminosity distance (for full definitions of the distance measures, see for example Appendix A of \cite{Czuchry:2023rbi}). On the data side, the observed distance modulus reads $\mu_{\rm data}=m - M$, where $m$ is the standardised apparent magnitude in the blue band, and $M$ is a fiducial absolute magnitude calibrated using the Cepheid host distances. In order to compare the model with $\Lambda$CDM, we also compute the $\chi^2$ values for our model. We form the measure $\Delta \mu$ depending on whether the SNeIa data points has an associated Cepheid host as 
\begin{equation}
    \Delta D_i = 
        \begin{cases}
            m_i - M - \mu^C_i, \quad i \in \text{Cepheid hosts} \\
            m_i - M - \mu_{\rm}, \quad \text{others},
        \end{cases}
\end{equation}
with the corresponding $\chi^2$ measure being
\begin{equation}
    \chi^2_{SneIa} = \left(\Delta \bm{D}\right)^T \left(C_{\rm tot}\right)^{-1} \left(\Delta \bm{D}\right),
\end{equation}
where $\Delta \bm{D} = \bm{D}_{\rm theory} - \bm{D}_{\rm data}$, and $C_{\rm tot}$ is a covariance matrix containing statistical and systematic uncertainties for both the SNeIa and Cepheids.

We also include measurements of the Hubble parameter from Passively-evolving Early-Type Galaxies (ETG), which have an old stellar population and thus a low star-formation rate. It is possible to reliably trace the spectral properties of ETG's along cosmic time (independent of the cosmological model), making ETG's a standardisable clock (they are also known as Cosmic Chronometers (CC)). For this purpose, we use a sample in the range $0 < z < 1.97$ \cite{Moresco:2012by,Moresco:2015cya,Gomez-Valent:2019lny}. In order to construct the $\chi^2$ for the CC's, we follow the same prescription as above and write
\begin{equation}
    \chi^2_{CC} = \left(\Delta\bm{H}\right)^T\left(C_{CC}\right)^{-1}\left(\Delta\bm{H}\right),
\end{equation}
where $\Delta \bm{H} = \bm{H}_{\rm theory} - \bm{H}_{\rm data}$, and $C_{CC}$ is a covariance matric containing statistical, sample-contamination, model dependence, and stellar metallicity uncertainties\footnote{It can be generated using the code \url{https://gitlab.com/mmoresco/CCcovariance}}.

We investigate the fit of our model to these two data sets using three sets of parameter values. Since we are using late-time data and are considering a flat Universe, the free parameters are $\{H_0, \, \Omega_m^0\}$. Since we are not performing a Bayesian likelihood analysis at this stage, we will fix the fiducial absolute magnitude $M=-19.5$, which is close to the canonical value. In this analysis, we use the average Hubble parameter \eqref{eq:hbar} in the definition of the distance measures. Here, we are varying only the standard cosmological parameters and do not consider the contribution to the energy densities of the scalar and gauge field, which are fixed by initial condition. As such, we are likely overestimating the value of $\Omega_m^0$ which in principle obtains contributions from the new scalar (depending on the equation of state), but this approach is enough to give an indication of the overall fit at late times. Since our numerical results in the previous sections indicate that the anisotropic effects are small at late times (low redshift), we pick the parameter values to lie close to, but slightly deviating from the $\Lambda$CDM values. As such, we are able to estimate the deviations induced by our model as well as its sensitivity to the parameter values.\footnote{Not all parameter sets are available to us, due to numerical limitations.}
\begin{figure}[h]
\centering
    \includegraphics[width=0.8\textwidth]{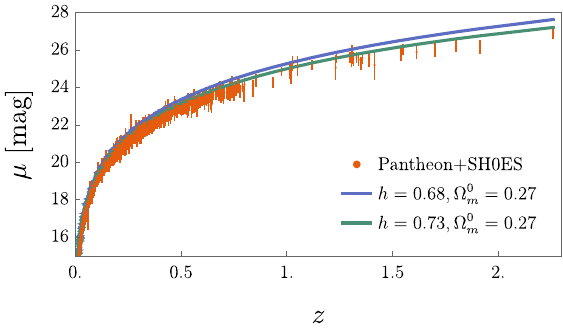}
    \caption{Distance modulus $\mu$ using TypeIa supernovae with Cepheid-distance calibration from the Pantheon+SH0ES dataset \cite{Brout:2022vxf} (orange) with $1\sigma$ error bars, together with the theoretical prediction from our model using for specific choices of parameter values. The parameter set $\{h=0.68,\Omega_m^0=0.27\}$ is not depicted here, as it overlaps with ($h=0.7,\Omega_m^0=0.30$) (at least graphically).}
    \label{fig:panth}
\end{figure}
\begin{figure}[h]
\centering
    \includegraphics[width=0.8\textwidth]{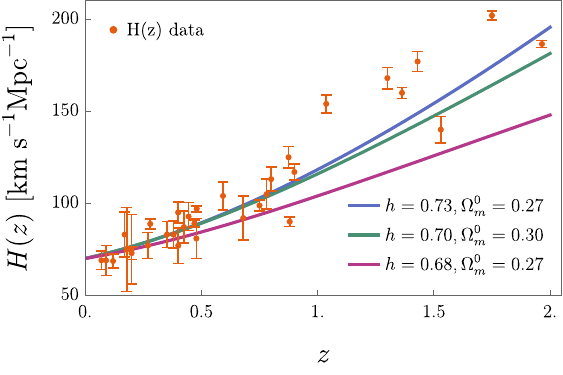}
    \caption{Hubble parameter measurements from Cosmic Chronometers (orange) with $1\sigma$ error bars \cite{Moresco:2012by,Moresco:2015cya}, together with the theoretical prediction from our model for specific choices of parameter values.}
    \label{fig:CC}
\end{figure}
Figures \ref{fig:panth} show the Pantheon+SH0ES data as a function of redshift (where $h=H_0/(100 \text{km}s^{-1}\text{Mpc}^{-1})$ is the dimensionless Hubble parameter). We observe that in general, raising the Hubble parameter from the CMB value ($h=0.68$) to the local Universe value ($h=0.73$) provides a better fit to the data. The same trend can be seen in Figure \ref{fig:CC} showing the data from Cosmic Chronometers, where the combination $h=0.73$, $\Omega_m^0=0.27$ provides the best fit to the data points. This also has implications for the age of the Universe as chosen in our analysis.

In order to compare with the $\Lambda$CDM model, we compute the $\chi^2$ statistic as described above, and we show the values in Table~\ref{tab:chi2}. We find that $\Lambda$CDM, (which for the present parameter values gives a $\chi^2 \sim 10^3$, the minimum $\chi^2$ being $1600$) fits the Pantheon+ SH0ES data in Figure~\ref{fig:panth} significantly better than our model (where we find $\chi^2\sim 10^5$), the difference in $\chi^2$ being around $10^5$. We also note that the best-fit parameter set (out of those considered) is not the same for our model as for $\Lambda CDM$; in fact, the best fit to our model ($\chi^2=2.57\cdot10^5$ for $\{h=0.73,\Omega_m^0=0.27\}$) is the \emph{worst} fit for $\Lambda$CDM ($\chi^2=3.68\cdot10^3$). The conclusion of this simple comparison is that even though $\Lambda$CDM provides a better fit overall, the best-fit parameters are likely to be different in our model.

For the CC data, the difference is much smaller, and for one parameter combination, the $\chi^2$ differ by a factor of $4$ as compared to $\Lambda CDM$. Table~\ref{tab:chi2} shows the $\chi^2$ values for the set of parameters chosen. For this dataset, our model shows very low sensitivity to the parameter set chosen, the best and worst $\chi^2$ differing only by $460.52-460.41=0.11$. In contrast, $\Lambda$CDM shows greater variability, with the $\chi^2$ ranging from $14.84$ (best), to $187.29$ (worst). Interestingly, the parameter set giving the lowest $\chi^2$ ($h=0.7,\Omega_m^0=0.3$) is the same for both models, which is in contrast to the Pantheon+ SH0ES case described above.
Overall, the fit to Pantheon+SH0ES data is worse for all parameter choices (not including the naturally higher $\chi^2$ due to the number of data points being higher), although it is possible that our model prefers a different value of $M$. A complete Bayesian inference analysis of the model (through which we will be able to place error bars and significance levels on the model parameters) is challenging, and will be presented in a forthcoming paper.
\begin{table}[h]
\begin{center}
\begin{tabular}{lcc}
    Pantheon+ SH0ES & & \\
    \hline\hline \\[-1mm]
    Parameter & $\chi^2$ (Our model) & $\chi^2$ ($\Lambda$CDM) \\
    \hline \\[-1mm]
    $h=0.7,\Omega_m^0=0.3$ & $2.61\cdot 10^5$ & $2.14\cdot 10^3$ \\
    $h=0.68,\Omega_m^0=0.27$ & $2.66\cdot 10^5$ & $1.60\cdot 10^3$\\
    $h=0.73,\Omega_m^0=0.27$ & $2.57\cdot 10^5$ & $3.68\cdot 10^3$ \\[3mm]
    CC & & \\
    \hline\hline \\[-1mm]
    Parameter & $\chi^2$ (Our model) & $\chi^2$ ($\Lambda$CDM) \\
    \hline \\[-1mm]
    $h=0.7,\Omega_m^0=0.3$ & $460.41$ & $14.84$ \\
    $h=0.68,\Omega_m^0=0.27$ & $460.52$ & $16.38$\\
    $h=0.73,\Omega_m^0=0.27$ & $460.35$ & $187.29$ \\
\end{tabular}
    \caption{$\chi^2$ for three different parameter combinations.}
\label{tab:chi2}
\end{center}
\end{table}

\section{Discussion \& Conclusions}\label{sec:disc}
In this paper we introduce an axion-electrodynamics model for the purpose of generation of cosmological anisotropies. Working with abelian gauge fields, we choose the components of the gauge field $A_\mu$ to be aligned with the Killing vectors of the Bianchi VII$_0$ metric, and we show that the field content satisfies the same isometries as Bianchi VII$_0$. We solve the resulting equations of motion numerically using a perturbative scheme where the zeroth order is the homogeneous isotropic limit; in this way, we obtain the canonical $\Lambda$CDM solutions at zeroth order, with anisotropic contributions appearing at first order. Thanks to the parametrisation of the gauge field, we obtain solutions to the anisotropic scale factors $\beta^{(1)}_i$ which are driven by the evolution of the gauge-field $A_\mu$, and by constructing the average Hubble parameter $\bar{H}$, we see that the deviation from $\Lambda$CDM is largest in the early universe, before relaxing down to the asymptotic $\Lambda$CDM fixed point. The magnitude of $\bar{H}$ is always smaller than $H_{\Lambda \rm CDM}$, and a negative slope at all times, which may have implications for the Hubble tension. Simultaneously, the isotropic scale factor exhibits approximately standard $\Lambda$CDM evolution throughout the history of the Universe, although the amplitude is consistently higher. Our solutions for the anisotropic scale factors exp$(\beta^{(1)}_1)$ and exp$(\beta^{(1)}_2)$ are very similar in amplitude, but not identical; this is a desirable feature, since cosmological anisotropies are expected to be small, and by evaluating exp$(\beta^{(1)}_1)$ and exp$(\beta^{(1)}_2)$ at the present time ($t_0=1/H_0$), we find that the anisotropic expansion is on the order of $10^{-7} - 10^{-8}$; by examining $\bar{H}$ in Figure~\ref{fig:Haverage}, we see that a large part of the anisotropies have decayed away at $t=5$ Gyr. The scalar field $\phi$  exhibits steep falloff in the early Universe and settles down to a small constant at late times, and we find similar behaviour in $\psi$ and $\sigma$, which parametrize the gauge field. A related model was studied in \cite{Watanabe:2009ct} and similar results were found, but as discussed in the Introduction, this is gauge-inequivalent to our model.

Taken together, these results indicate that most non-trivial effects will be contained to the early universe. Whilst this does safeguard late-time evolution against large anisotropic effects, this is not necessarily desirable, since early-Universe processes (inflation, BBN, recombination etc) are very sensitive to the field content and initial conditions; in particular, early-Universe observables such as the sound horizon may be modified in the presence of anisotropies, in an analogous way to that of early dark energy \cite{Kamionkowski:2022pkx}.However, this lies beyond the scope of the present work. For studies regarding anisotropies in the inflationary era, see for example \cite{Watanabe:2009ct,Dulaney:2010sq,Gumrukcuoglu:2007bx,Gumrukcuoglu:2010yc,Pitrou:2008gk}.

In ~\ref{app:pertexp} we find that the perfect-fluid part of the total stress-energy tensor receives anisotropic corrections perturbatively, both in the energy density and in the pressure. We also find off-diagonal components to the stress-energy tensor, which act as constraint equations, as was also studied in~\cite{Cho:2022rgs}. The anisotropic part of the energy density has been studied as \emph{anisotropic dark energy}, for example in \cite{Koivisto:2005mm} and \cite{Koivisto:2008ig}, although at the background level. There are also interesting connections to the quadrupole anomaly in the CMB \cite{Rodrigues:2007ny}. 

The most important result of this work is the   generation of cosmological anisotropies; we have shown that it is possible to find solutions which closely resemble those of $\Lambda$CDM at zeroth order, whilst containing a small degree of anisotropic correction at order $\epsilon$. An important note is that we are likely overestimating the magnitude of the dark-energy density $\Omega_\Lambda$: since the extra field content $\{\phi(t),\psi(t),\sigma(t),\beta_1(t),\beta_2(t)$ can be interpreted as dynamical dark energy, the total dark-energy density should read $\Omega_{\rm DE}=\Omega_{\Lambda}+\Omega_{\phi} + \hdots$, but because of the small scales of the anisotropies and the field $\phi(t)$, this would be a very small correction\footnote{For a discussion of the current observational status of dynamical dark energy, see \cite{SolaPeracaula:2018wwm}.}.

The observational status of cosmological anisotropy is rapidly evolving, with some groups claiming very strong results, such as anisotropic acceleration (anomalous bulk flow) in the direction of the CMB dipole at $3.9\sigma$ significance~\cite{Colin:2019opb} and a $3\sigma$ hemispherical power asymmetry in the Hubble constant, also aligned with the CMB dipole\footnote{A possible solution to the hemispherical power asymmetry was recently proposed in \cite{Kumar:2022zff}.}~\cite{Luongo:2021nqh}. Together with probes such as fine structure-constant variation and preferred directions in the CMB results in compelling evidence that the cosmological standard model needs revision, and we have provided a mechanism through which such preferred directions can arise dynamically from a well-motivated field theory. This is of course not the only model which can generate cosmological anisotropies; in particular, models exhibiting spacetime-symmetry breaking are known to contain preferred directions in the form of timelike vector fields. For example Ho\v{r}ava-Lifshitz gravity \cite{Horava:2009uw} Einstein-Aether theory \cite{Gasperini:1987nq}, and  bumblebee gravity~\cite{Maluf:2021lwh}, all of which have received significant attention in recent years, contain preferred frames of reference. On the other hand, spacetime-symmetry breaking in gravity has been tightly constrained using the Standard-Model Extension effective field theory, restricting the available parameter space for all spacetime-symmetry breaking models~\cite{Kostelecky:2008ts}. Our construction has the advantage of keeping these well-tested spacetime symmetries intact, and instead postulating the existence of the fields $\psi(t)$ and $A_\mu(t)$, and in this sense, it can be considered a scalar-vector model. In this paper, we presented an estimate of the fit of the average Hubble parameter of our model with late-time cosmological data for certain parameter values and contrasted it with the fit using $\Lambda$CDM. Overall, the standard cosmological model is a better fit to late-time probes.

It is worthwhile to mention \cite{Krishnan:2022qbv}, which have  partial overlap, and of course are compatible with some of the   results and statements presented in this paper. However it is important to note that the authors of \cite{Krishnan:2022qbv} considered so called flowing dark-energy cosmology, where the tilt parameter is non-zero at late times (for details see Section V of \cite{Krishnan:2022qbv}). This is in contrast to the model in the present paper, where all anisotropies decay to a homogeneous and isotropic fixed point, in keeping with the cosmic no-hair theorem~\cite{Wald:1983ky}. Also, even though FLRW is stable in our setup, the possibility of a tilt instability in the FLRW geometry which could potentially evade detection through the cosmic no-hair theorem was raised
in \cite{Krishnan:2022uar}.

A natural extensions and applications of this work would be to consider an $SU(2)$ gauge field, as was done in the context of cosmic birefringence in \cite{Ishiwata:2021yne}, as well as computing imprints of anisotropy on the CMB, by introducing angular dependence of the metric functions. All of these applications, as well as parameter constraints on the present model by means of cosmological data and a Markov-Chain Monte Carlo (MCMC) algorithm are forthcoming \cite{Lee:2023azx,inprogress}.

\section*{Acknowledgements}
We thank Stephen Appleby, Eoin Ó Colgáin, Jeong-Hyuck Park, and Inyong Cho for discussions and comments on the draft. BHL thanks APCTP for the hospitality during his visit, while part of this work has been done. BHL, WL, HL, and  NAN were supported by the Basic Science Research Program through the National Research Foundation of Korea (NRF) funded by the Ministry of Education, Science and Technology (BHL, HL, NAN: NRF-2020R1A6A1A03047877, BHL, HL: NRF-2020R1F1A1075472, WL: NRF-2022R1IA1A01067336, NAN: NRF-2020R1A6A1A03047877, \\ HL: NRF-2016R1D1A1B01015196). NAN is also grateful for financial support by CNES. The work of ST was supported by Mid-career Researcher Program through the National Research Foundation of Korea grant No. NRF-2021R1A2B5B02002603.

\appendix
\section{General Formalism of Bianchi Metrics}\label{app:biagen}
The most general metric of the Bianchi geometries can be written as \cite{Misner:1973prb}
\begin{align}\label{genmet}
d s^2=-N(t)^2 d t^2+e^{2 \alpha(t)} e^{2 \beta_{i j}(t)} \omega^i \omega^j,
\end{align}
where $N(t)$ is the lapse function, $\omega^i$ are  1-forms, $e^{2 \alpha(t)}$ is the scale factor of the universe and $\beta_{i j}$ determines the anisotropic parameters.
In this general Bianchi model, the shift vector is not stipulated in the metric, and the lapse function can consequently be a dynamical variable; however, in the flat-space limit this can be gauged away and we can safely set this lapse function to a constant \cite{Lorenz:1980kq}.
 In Eq.~\eqref{genmet}, $\beta_{i j}$ determines the anisotropic parameters, which can in principle be a general matrix with non-diagonal entries. However we can work in a diagonal basis where  $\beta_{+}(t)$ and $\beta_{-}(t)$ are given as follows
\begin{equation}
\beta_{i j}=\left(\begin{array}{ccc}
\beta_{+}+\sqrt{3} \beta_{-} & 0 & 0 \\
0 & \beta_{+}-\sqrt{3} \beta_{-} & 0 \\
0 & 0 & -2 \beta_{+}
\end{array}\right)
\end{equation}
In this paper we consider the Bianchi metric in such a diagonal basis of $\beta_{i j}$, and we identify
\begin{align}
    \beta_{+}+\sqrt{3} \beta_{-}=\beta_1, \beta_{+}-\sqrt{3} \beta_{-}=\beta_2, -2 \beta_{+}=\beta_3,
\end{align}
from which we obtain the most general Bianchi I metric ansatz in Cartesian coordinates as
\begin{equation}\label{eq:metric}
    ds^2 = -dt^2 + e^{2\alpha(t)}\left(e^{2\beta_1(t)}dx_1^2 + e^{2\beta_2(t)}dx_2^2 + e^{2\beta_3(t)}dx_3^2\right),
\end{equation}
where $\alpha(t)$ and $\beta_i(t)$ are the isotropic and anisotropic scale factors, respectively. The factor two has been introduced so that the isotropic scale factor matches its FLRW equivalent, i.e. $a(t)=\text{exp}(\alpha(t))$, and $\dot{a}/a = \dot{\alpha}$.

In this Appendix we have presented a general treatment of the Bianchi type, but in the
rest of the paper (and \ref{klsymgf}) we specify our geometry to Bianchi type VII$_0$, which is suitable for our purposes and keeps the equations tractable. In a nutshell, Bianchi-type geometries are classified by their Killing vector fields, and in our present work we only consider Bianchi Type VII$_0$. For an exhaustive Bianchi classification, we refer the reader to \cite{Papadopoulos:2011rk}.

\section{Killing Symmetry of the Gauge Fields}\label{klsymgf}
In this appendix we explicitly show that the $U(1)$ gauge field under consideration has the Killing symmetries of the metric.

Let us start with the metric in the Bianchi I metric as follows
\begin{equation}\label{b7}
\begin{aligned}
    ds^2 = -dt^2 + e^{2\alpha(t)+2\beta_1(t)}dx_1^2& + e^{2\alpha(t)+2\beta_2(t)}dx_2^2 +e^{2\alpha(t)+2\beta_3(t)}dx_3^2.
\end{aligned}
\end{equation}
We have three Killing vectors associated with the \eqref{b7}, 
$$K_i=\partial_i,$$
with $i=x_i's$. The Killing vectors satisfies the following condition $$\mathcal{L}_{K_i} g^{\mu\nu}=0$$. Here we will use a convenient notation for $\omega_i\wedge \omega_j $ just by $\omega_i\omega_j$ with the property that $$\omega_i\omega_j=-\omega_j\omega_i.$$

Let us write the 2-form fluxes in the most general form as 
\begin{equation}
\begin{aligned}
    F=&\left(f_1 dtdx_1+f_2 dt dx_2+f_3dt dx_3\right)+\left(g_1dx_1dx_2+g_2dx_2dx_3+g_3dx_3dx_1\right),
\end{aligned}
\end{equation}
where the $f_i$'s and $g_i$'s can be arbitrary functions of $(t,x_i)$. The equation of motion of the 2-form fields are given  by $$d F=0,$$ which gives us the following constraint equations
\begin{eqnarray}\label{eq2fr}
    \left(\partial_2f_1-\partial_1f_2+\partial_tg_1\right)&=&0,\nonumber\\
    \left(\partial_3f_1-\partial_1f_3-\partial_tg_3\right)&=&0,\nonumber\\
    \left(\partial_3f_2-\partial_2f_3+\partial_tg_2\right)&=&0,\nonumber\\
    \left(\partial_1g_2+\partial_2g_3+\partial_3g_1\right)&=&0.
\end{eqnarray}
The Killing equation is given by the following equation
\begin{equation}
	\mathcal{L}_{K_i}F=d(iK_i F)+iK_i(dF).
\end{equation}
Now if the 2-form field has the same Killing symmetry, then $$\mathcal{L}_{K_i}F=0$$ identically, which after a few trivial algebraic manipulation gives the following constraint equations,
\begin{equation}\label{kl2fr}
\begin{aligned}
    -(\partial_1 f_1 d_{01}&+\partial_1f_2 d_{02}+\partial_1 f_3 d_{03})-\left(\partial_1g_1d_{12}+\partial_1g_3d_{31}+\partial_1g_2d_{23}\right)=0,\\
    -(\partial_2 f_1 d_{01}&+\partial_2f_2 d_{02}+\partial_2 f_3 d_{03})-\left(\partial_2g_1d_{12}+\partial_2g_3d_{31}+\partial_2g_2d_{23}\right)=0,\\
    -(\partial_3 f_1 d_{01}&+\partial_3f_2 d_{02}+\partial_3 f_3 d_{03})-\left(\partial_3g_1d_{12}+\partial_3g_3d_{31}+\partial_3g_2d_{23}\right)=0.
\end{aligned}
\end{equation}
Where we have defined the volume form as 
$$d_{ij}=dx_idx_j.$$
Simultaneously satisfying \eqref{eq2fr} and \eqref{kl2fr} gives us the following relations,
\begin{align}
    \partial_j f_i=0,\qquad \partial_jg_i=0,\qquad i,j=1,2,3
\end{align}
which implies that $f_i$ and $g_i$ are functions of only time $t$.
 
A similar analysis can be done for the 1-form gauge fields $A_i$. The 1-form gauge field can be written as 
\begin{equation}
A=a_0dt+b_i dx_i,
\end{equation}
where again $a_0$ and $b_i$ can be arbitrary functions of $(t,x_i)$.
The fluxes can be computed as $F=dA$ and the equation of motion is trivially satisfied, $$dF=0.$$
Lets us write the Killing equation for the 1-form $A$
$$\mathcal{L}_{K_{i}}(A)=d(iK_i A)+iK_i(dA)$$ and the algebra can be easily worked out 
\begin{eqnarray}\label{klggf}
\mathcal{L}_{Kj}A=\partial_j a_0 dt+\partial_jb_i dx_i,\qquad i,j=1,2,3.
\end{eqnarray}
Satisfying the Killing equation leads to
\begin{eqnarray}\label{1frklcon}
 \partial_j a_0 =0\qquad \partial_jb_i =0,\qquad i,j=1,2,3,
\end{eqnarray}
which implies that $a_0$ and $b_i$ can at most be a function of the time $t$ only.
So far we have worked with the most general Killing symmetry of Bianchi $I$ type; however, we can similarly generalize to the metric ansatz we have implemented in our main text,  Bianchi VII$_0$ with $\beta_2=\beta_3$ in \eqref{b7} which will have the following Killing vectors 
\begin{equation}
k_i=\partial_i,\qquad i=2,3\qquad \text{and}\qquad k_1=\partial_{1}-(x_2\partial_3-x_3\partial_2)
\end{equation}
 Note here that once we consider more general metric ansatz, which depends on angular direction, we have different sets of Killing vectors, and the $1-$form field strengths can depend on these variable. We defer this analysis for our forthcoming work \cite{inprogress}.

 \section{Metric Gauge Choice}\label{app:metggc}
In this section we give some relevant arguments for the gauge choice of the metric ansatz we use in this paper. The isotropic scale factor $\alpha$ can also be expanded in a perturbative expansion as follows $$\alpha(t)=\alpha^{(0)}+\epsilon ~\alpha^{(1)}+\cdots.$$ However the gauge degrees of freedom of the metric  allows us to set $\alpha^{(1)}=0$ and set all the contribution of the anisotropy in the $\beta's$.
 In such gauge choice with  $\alpha^{(1)}(t) \neq 0$, there will be non-vanishing contribution coming at the first order in the isotropic scale factor. In such set up the matter sector will non-trivially back-react on the metric and we will have corrections to the isotropic scale factors. For example if we study Quintessence model with such metric ansatz then the Quintessence (scalar) fields would back-react at the first order in the isotropic scale factors. Thus such first order correction to the isotropic scale factors would in principle differ significantly from the base $\Lambda$CDM Quintessence models.
In the present work we mostly focus on the anisotropic scale factors, which are generated at the first order due to the back-reaction of the matter sector on the metric. So one natural choice would be absorb the first order correction to the anisotropic scale factors in the $\beta_i$'s or equivalently we can set the  $\alpha^{(1)}(t)=0.$
\par
\vspace{5mm}
{\bf Defining a comoving-volume preserving coordinate:}\\
Here, we outline an argument for the introduction of the comoving-volume conserving coordinate, and we show that this is fully compatible with the computation presented in the paper.
The metric used in our analysis is of the form 
\begin{equation}\label{eq:metricB1app}
    ds^2 = -dt^2 + e^{2\alpha(t)}\left(e^{2\beta_1(t)}dx_1^2 + e^{2\beta_2(t)}dx_2^2 + e^{2\beta_3(t)}dx_3^2\right),
\end{equation}
Now, we can redefine the isotropic and anisotropic scale factor as follows
\begin{equation}\label{scfredef}
  \begin{aligned}
    \tilde{\alpha}(t) &= \alpha(t)+f(t)+c,\\
    \tilde{\beta}_i(t) &= \beta(t)_i-f(t)-c,
  \end{aligned}  
\end{equation}
where $f(t)$ is some function of time which we will derive below and $c$ is an arbitrary constant. It is very easy to check that with the above redefinition \eqref{scfredef} the metric is invariant,
\begin{equation}\label{eq:metricB1red}
    ds^2 = -dt^2 + e^{2\tilde{\alpha}(t)}\left(e^{2\tilde{\beta}_1(t)}dx_1^2 + e^{2\tilde{\beta}_2(t)}dx_2^2 + e^{2\tilde{\beta}_3(t)}dx_3^2\right).
\end{equation}
Now imposing the constraint 
\begin{equation}
\sum_i\dot{\tilde{\beta}}_i=0.
\end{equation}
It is easy to see that
\begin{equation}
\dot{f}(t)=\tfrac{1}{3}\sum_i \dot{\beta}(t)_i.
\end{equation}
With such a redefinition the isotropic scale factor, viz, $\alpha(t)$ is redefined to include contribution from the anisotropic scale factors, $\beta_i$, and the properties of $\tilde{\alpha}(t)$ are follows

\begin{itemize}
    \item $\tilde{\alpha}(t)$ should flow to the isotropic fixed point asymptotically, and hence the $\beta_i$'s should vanish at the isotropic fixed point. Here, we have explicitly included and showed the asymptotic vanishing of the anisotropic scale factors at the isotropic fixed points. 
    \item The constant $c$ can be reabsorbed by a coordinate redefinition, which is equivalent to setting it to zero.
    \item $\tilde{\beta}_i$ should flow to the attractor fixed point asymptotically, so with the above redefinition we have a direct parallel between our analysis and the comoving-volume conserving coordinates. We would however like to to emphasize that with constant comoving volume gauge, the isotropic scale factor $\alpha$ will be different from what we expect from the FLRW scale factor.
\end{itemize}
 Defining a comoving-volume preserving metric puts a constraint on the anisotropic scale factors: as noted in the main text , the first order of the first Friedmann equation, (the $tt$ component of the Einstein/Friedmann equation) constrains the anisotropic scale factors.

\section{Choice of initial conditions}\label{app:const}
In this appendix we write the constraints coming from the Einstein equations and the other field equations which we implement in the numerical solutions. A priori even though the functions $\alpha,\beta_1, \beta_2$ and their derivatives seems to be independent, however differential  equations sets some constraints on there functions.
Below we write out the constraints equations and we write our choice of boundary values in Table~\ref{tab:initconds}.

\begin{enumerate}
    \item The first Friedmann equation at the zeroth order sets the constraint on $\dot{\alpha(t)}$ once we have chosen the initial value of $\alpha(t).$
    \item Similarly, the first order (${\cal O}(\epsilon)$) of the first Friedmann equation sets a constraint between the $\beta_i$'s: once the initial values of  $\beta_1, \beta_2$ and $\dot{\beta_1}$ are fixed the  value of $\dot{\beta}^{(1)}_2(t)$ is constrained by the other equations.
    
    \item The zeroth order off-diagonal part of the Einstein equations fixes the initial value of $\dot{\psi}^{(0)}(t)$ once we specify the $\psi^{(0)}(t)$ and $\alpha^{(0)}(t):$
    $$\dot{\psi}^{(0)}(t_f)=-\dot{\alpha}^{(0)}(t_f) \psi^{(0)}(t_f).$$
\end{enumerate}

\section{Perturbative Expansions}\label{app:pertexp}

In this section, we reintroduce the coupling $\kappa$ for completeness.
The $\mu=1$ component of the vector equations read

\begin{align}
\nonumber 0 \;&=\;3 \dot{\alpha}^{(0)} \dot{\psi}^{(0)}+\psi^{(0)} \left(\ddot{\alpha}^{(0)}+2 (\dot{\alpha}^{(0)})^2\right)+\ddot{\psi}^{(0)} +\epsilon\Big[\psi^{(1)} \ddot{\alpha}^{(0)}+2 \psi^{(0)} \dot{\alpha}^{(0)} \dot{\beta}^{(1)}_1 +2 \psi^{(0)} \dot{\alpha}^{(0)} \dot{\beta}^{(1)}_2\\ \nonumber &-6 \psi^{(0)} \dot{\alpha}^{(0)} \dot{\sigma}^{(1)}  +3 \dot{\alpha}^{(0)} \dot{\psi}^{(1)}+2 \psi^{(1)} (\dot{\alpha}^{(0)})^2 +\beta^{(1)}_1 \Big(3 \dot{\alpha}^{(0)} \dot{\psi}^{(0)}+\psi^{(0)} \left(\ddot{\alpha}^{(0)}+2 (\dot{\alpha}^{(0)})^2\right)+\ddot{\psi}^{(0)}\Big)\\ \nonumber & -2 \sigma^{(1)} \Big(3 \dot{\alpha}^{(0)} \dot{\psi}^{(0)}+\psi^{(0)} \left(\ddot{\alpha}^{(0)}+2 (\dot{\alpha}^{(0)})^2\right)+\ddot{\psi}^{(0)}\Big)+\psi^{(0)} \ddot{\beta}^{(1)}_1+\dot{\beta}^{(1)}_1 \dot{\psi}^{(0)}+2 \dot{\beta}^{(1)}_2 \dot{\psi}^{(0)}\\& -2 \psi^{(0)} \ddot{\sigma}^{(1)}-4 \dot{\sigma}^{(1)} \dot{\psi}^{(0)}+\ddot{\psi}^{(1)}\Big]
\end{align}

and the $\mu=2,3$ component is
\begin{align}
\nonumber 0 \;&=\; 3 \dot{\alpha}^{(0)} \dot{\psi}^{(0)}+\psi^{(0)} \left(\ddot{\alpha}^{(0)}+2 (\dot{\alpha}^{(0)})^2\right)+\ddot{\psi}^{(0)}+\epsilon  \Big[(\psi^{(1)} \ddot{\alpha}^{(0)}+\psi^{(0)} \dot{\alpha}^{(0)} \dot{\beta}^{(1)}_1+3 \psi^{(0)} \dot{\alpha}^{(0)} \dot{\beta}^{(1)}_2\\ \nonumber &+3 \psi^{(0)} \dot{\alpha}^{(0)} \dot{\sigma}^{(1)}+3 \dot{\alpha}^{(0)} \dot{\psi}^{(1)} +2 \psi^{(1)} (\dot{\alpha}^{(0)})^2+\beta^{(1)}_2 \left(3 \dot{\alpha}^{(0)} \dot{\psi}^{(0)}+\psi^{(0)} \left(\ddot{\alpha}^{(0)}+2 (\dot{\alpha}^{(0)})^2\right)+\ddot{\psi}^{(0)}\right)\\ \nonumber &+\sigma^{(1)} \left(3 \dot{\alpha}^{(0)} \dot{\psi}^{(0)}+\psi^{(0)} \left(\ddot{\alpha}^{(0)}+2 (\dot{\alpha}^{(0)})^2\right)+\ddot{\psi}^{(0)}\right) +\dot{\beta}^{(1)}_1 \dot{\psi}^{(0)}+\psi^{(0)} \ddot{\beta}^{(1)}_2+2 \dot{\beta}^{(1)}_2 \dot{\psi}^{(0)}+\psi^{(0)} \ddot{\sigma}^{(1)}\\ &+2 \dot{\sigma}^{(1)} \dot{\psi}^{(0)}+\ddot{\psi}^{(1)}\Big],
\end{align}

The $\mu=\nu=0$ component of the Einstein equations (the first Friedmann equation) read
\begin{align}
\nonumber & 3 (\dot{\alpha}^{(0)})^2+ \epsilon  \Big[2 \dot{\alpha}^{(0)} \dot{\beta}^{(1)}_1+4 \dot{\alpha}^{(0)} \dot{\beta}^{(1)}_2\Big] = \frac{\kappa}{2} \left[3 \left(\psi^{(0)} \dot{\alpha}^{(0)}+\dot{\psi}^{(0)}\right)^2 +2 \rho^{\rm PF} +2 V(\phi^{(0)})+(\dot{\phi}^{(0)})^2\right] \\ \nonumber & -\epsilon \kappa    \Big\{-(\psi^{(0)})^2 \dot{\alpha}^{(0)} \left(\dot{\beta}^{(1)}_1+2 \dot{\beta}^{(1)}_2\right)-\psi^{(0)} \Big[3 \dot{\alpha}^{(0)} \left(\psi^{(1)} \dot{\alpha}^{(0)}+\dot{\psi}^{(1)}\right) +\dot{\psi}^{(0)} \left(\dot{\beta}^{(1)}_1+2 \dot{\beta}^{(1)}_2\right)\Big] \\ & -3 \psi^{(1)} \dot{\alpha}^{(0)} \dot{\psi}^{(0)}-\phi^{(1)} V'(\phi^{(0)})-3 \dot{\psi}^{(0)} \dot{\psi}^{(1)}-\dot{\phi}^{(0)} \dot{\phi}^{(1)}\Big\},
\end{align}
the $\mu=\nu=1$ component is
\begin{align}
  \nonumber & 2 \ddot{\alpha}^{(0)} + 3 (\dot{\alpha}^{(0)})^2+\epsilon  \left[4 \beta^{(1)}_1 \ddot{\alpha}^{(0)}+6 \beta^{(1)}_1 (\dot{\alpha}^{(0)})^2+6 \dot{\alpha}^{(0)} \dot{\beta}^{(1)}_2+2 \ddot{\beta}^{(1)}_2\right]=-\frac{\kappa}{2}  \Big[(\psi^{(0)})^2 (\dot{\alpha}^{(0)})^2\\ \nonumber &+2 \psi^{(0)} \dot{\alpha}^{(0)} \dot{\psi}^{(0)} +2 p^{\rm PF}-2 V(\phi^{(0)})+(\dot{\psi}^{(0)})^2+(\dot{\phi}^{(0)})^2\Big]-\epsilon \kappa \Big[- (\psi^{(0)})^2 \dot{\alpha}^{(0)} \dot{\beta}^{(1)}_1+ \beta^{(1)}_1 (\psi^{(0)})^2 (\dot{\alpha}^{(0)})^2\\ \nonumber &+2 \beta^{(1)}_1 \psi^{(0)} \dot{\alpha}^{(0)} \dot{\psi}^{(0)}+2(\psi^{(0)})^2 \dot{\alpha}^{(0)} \dot{\beta}^{(1)}_2+4 (\psi^{(0)})^2 \dot{\alpha}^{(0)} \dot{\sigma}^{(1)}+4 \sigma^{(1)} (\psi^{(0)})^2 (\dot{\alpha}^{(0)})^2  +8 \sigma^{(1)} \psi^{(0)} \dot{\alpha}^{(0)} \dot{\psi}^{(0)}\\ \nonumber &+ \psi^{(1)} \dot{\alpha}^{(0)} \dot{\psi}^{(0)}+ \psi^{(0)} \dot{\alpha}^{(0)} \dot{\psi}^{(1)}+ \psi^{(0)} \psi^{(1)} (\dot{\alpha}^{(0)})^2 - \psi^{(0)} \dot{\beta}^{(1)}_1 \dot{\psi}^{(0)}+ \beta^{(1)}_1 (\dot{\psi}^{(0)})^2 + \beta^{(1)}_1 (\dot{\phi}^{(0)})^2\\ \nonumber &+2 \psi^{(0)} \dot{\beta}^{(1)}_2 \dot{\psi}^{(0)}+2 \beta^{(1)}_1 p^{\rm PF}+4 \psi^{(0)} \dot{\sigma}^{(1)} \dot{\psi}^{(0)}+4 \sigma^{(1)} (\dot{\psi}^{(0)})^2- \phi^{(1)} V'(\phi^{(0)})-2 \beta^{(1)}_1 V(\phi^{(0)}) \\ & + \dot{\psi}^{(0)} \dot{\psi}^{(1)}+ \dot{\phi}^{(0)} \dot{\phi}^{(1)}\Big].
\end{align}
$\mu=\nu=2,3$ are equal and read
\begin{align}
       \nonumber & 4 \ddot{\alpha}^{(0)}+6 (\dot{\alpha}^{(0)})^2+\epsilon  \left[8 \beta^{(1)}_2 \ddot{\alpha}^{(0)}+6 \dot{\alpha}^{(0)} \dot{\beta}^{(1)}_1+6 \dot{\alpha}^{(0)} \dot{\beta}^{(1)}_2+12 \beta^{(1)}_2 (\dot{\alpha}^{(0)})^2+2 \ddot{\beta}^{(1)}_1+2 \ddot{\beta}^{(1)}_2\right]= \\ \nonumber & -\kappa  \Big((\psi^{(0)})^2 (\dot{\alpha}^{(0)})^2  +2 \psi^{(0)} \dot{\alpha}^{(0)} \dot{\psi}^{(0)}+2 p^{\rm PF}-2 V(\phi^{(0)})+(\dot{\psi}^{(0)})^2+(\dot{\phi}^{(0)})^2\Big)-\epsilon \kappa \Big[2 (\psi^{(0)})^2 \dot{\alpha}^{(0)} \dot{\beta}^{(1)}_1\\ \nonumber &+2 \beta^{(1)}_2 (\psi^{(0)})^2 (\dot{\alpha}^{(0)})^2  +4 \beta^{(1)}_2 \psi^{(0)} \dot{\alpha}^{(0)} \dot{\psi}^{(0)}-4 (\psi^{(0)})^2 \dot{\alpha}^{(0)} \dot{\sigma}^{(1)} -4 \sigma^{(1)} (\psi^{(0)})^2 (\dot{\alpha}^{(0)})^2\\ \nonumber &-8 \sigma^{(1)} \psi^{(0)} \dot{\alpha}^{(0)} \dot{\psi}^{(0)}+2 \psi^{(1)} \dot{\alpha}^{(0)} \dot{\psi}^{(0)}+2 \psi^{(0)} \dot{\alpha}^{(0)} \dot{\psi}^{(1)} +2 \psi^{(0)} \psi^{(1)} (\dot{\alpha}^{(0)})^2 +2 \psi^{(0)} \dot{\beta}^{(1)}_1 \dot{\psi}^{(0)}\\ \nonumber &+2 \beta^{(1)}_2 (\dot{\psi}^{(0)})^2+2 \beta^{(1)}_2 (\dot{\phi}^{(0)})^2+4 \beta^{(1)}_2 p^{\rm PF}-4 \psi^{(0)} \dot{\sigma}^{(1)} \dot{\psi}^{(0)} \\ & -4 \sigma^{(1)} (\dot{\psi}^{(0)})^2-2 \phi^{(1)} V'(\phi^{(0)})-4 \beta^{(1)}_2 V(\phi^{(0)})+2 \dot{\psi}^{(0)} \dot{\psi}^{(1)}+2 \dot{\phi}^{(0)} \dot{\phi}^{(1)}\Big].
\end{align}
For the magnetic components, we obtain for $\mu=1, \nu=2$
\begin{align}
    \nonumber 0\;&=\;\left(\psi^{(0)} \dot{\alpha}^{(0)}+\dot{\psi}^{(0)}\right)^2+\epsilon \left(\psi^{(0)} \dot{\alpha}^{(0)}+\dot{\psi}^{(0)}\right) \Big[\beta^{(1)}_1 \psi^{(0)} \dot{\alpha}^{(0)}+\beta^{(1)}_2 \psi^{(0)} \dot{\alpha}^{(0)}-\sigma^{(1)} \psi^{(0)} \dot{\alpha}^{(0)}\\&+2 \psi^{(1)} \dot{\alpha}^{(0)} +\psi^{(0)} \dot{\beta}^{(1)}_1+\dot{\psi}^{(0)} (\beta^{(1)}_1+\beta^{(1)}_2-\sigma^{(1)})+\psi^{(0)} \dot{\beta}^{(1)}_2-\psi^{(0)} \dot{\sigma}^{(1)}+2 \dot{\psi}^{(1)}\Big],
\end{align}
and for $\mu=2, \nu=3$ we obtain
\begin{align}
    \nonumber 0\;&=\;\left(\psi^{(0)} \dot{\alpha}^{(0)}+\dot{\psi}^{(0)}\right)^2+\epsilon \left(\psi^{(0)} \dot{\alpha}^{(0)}+\dot{\psi}^{(0)}\right) \Big[\beta^{(1)}_2 \left(\psi^{(0)} \dot{\alpha}^{(0)}+\dot{\psi}^{(0)}\right)+\sigma^{(1)} \Big(\psi^{(0)} \dot{\alpha}^{(0)}\\ &+\dot{\psi}^{(0)}\Big)+\psi^{(1)} \dot{\alpha}^{(0)}+\psi^{(0)} \left(\dot{\beta}^{(1)}_2+\dot{\sigma}^{(1)}\right)+\dot{\psi}^{(1)}\Big].
\end{align}
The electric components are zero, since $\partial_i\phi(t)=0$.

\vspace{5mm}
\noindent\underline{\emph{Expansion of the perfect fluid stress-energy tensor}}\\

The continuity equation for the perfect fluid reads
\begin{equation}
    \dot{\rho}_i+3\bar{H}(1+w_i)\rho_i = 0,
\end{equation}
where $\bar{H}$ is the average Hubble parameter \eqref{eq:hbar}. The equation solves as
\begin{equation}
    \rho_i = \rho_i^0e^{-\left(1+w_i\right)\left[3\alpha^{(0)}+\epsilon\left(\beta^{(1)}_1+2\beta^{(1)}_2\right)\right]}.
\end{equation} 
From this we see that the perfect fluid stress-energy tensor will have corrections at the perturbative level, and we can write the components of $T_{\mu\nu}^{\rm PF}$ as
\begin{align}
    \nonumber T_{00}^{\rm PF} &= 3H_0^2\left(\Omega_r^0 e^{-4\alpha^{(0)}}+\Omega_m^0 e^{-3\alpha^{(0)}}+\Omega_\Lambda\right) \\&+ \epsilon \Big[ -H_0^2 \left(\beta^{(1)}_1+2\beta^{(1)}_2\right)\Big(4\Omega_r^0e^{-4\alpha^{(0)}}+3\Omega_m^0e^{-3\alpha^{(0)}}\Big) \Big] \\
    \nonumber T_{ii}^{\rm PF} &= 3H_0^2\left(\frac{1}{3}\Omega_r^0e^{-2\alpha^{(0)}}-\Omega_\Lambda^0e^{2\alpha^{(0)}}\right) + \epsilon \left\{ -H_0^2 \left[\frac{4}{3}\left(\beta^{(1)}_1+2\beta^{(1)}_2\right)\Omega_r^0e^{-2\alpha^{(0)}} \right.\right.\\& \left.\left. \nonumber-6\beta_i^{(1)}\left(\frac{1}{3}\Omega_r^0 e^{-2\alpha^{(0)}} - \Omega_\Lambda^0e^{2\alpha^{(0)}}\right)\right]\right\},
\end{align}
to first order in $\epsilon$, where we obtain the zeroth-order (isotropic) form as in Eq.~\eqref{eq:PF}. From this, we form the pressure $p_i$ as
\begin{equation}
\begin{aligned}
     p_i &= 3H_0^2\left(\frac{1}{3}\Omega_r^0e^{-4\alpha^{(0)}}-\Omega_\Lambda^0\right) \\&\quad \quad +\epsilon\left[-\frac{4}{3}H_0^2 \left(\beta^{(1)}_1+2\beta^{(1)}_2\right) \Omega_r^0 e^{-4\alpha^{(0)}}\right],
\end{aligned}
\end{equation}
by multiplying by $g^{ii}$ to first order in $\epsilon$ As such, the perfect fluid received \emph{anisotropic} corrections to the pressure in the presence of a radiation term. Note that we have taken $\Omega_r^0 = 0$ in this paper, and that this term would only be significant at very early times.

\newpage
\bibliographystyle{unsrt}
\bibliography{apssamp}
\end{document}